\documentclass[twocolumn,superscriptaddress,aps,preprintnumbers,amsmath,amssymb,prd,nofootinbib]{revtex4-2}

\usepackage{graphicx}
\usepackage{subcaption}
\usepackage{epsfig}
\usepackage{epstopdf}
\usepackage{caption}
\usepackage{url}
\usepackage{xcolor}
\usepackage[T1]{fontenc}
\usepackage{float}
\usepackage[bitstream-charter]{mathdesign}
\usepackage[nointegrals]{wasysym} 

\DeclareSymbolFont{myletters}{OML}{ztmcm}{m}{it}
\DeclareMathSymbol{\uplambda}{\mathord}{myletters}{"15}
\DeclareMathSymbol{\upxi}{\mathord}{myletters}{"18}

\usepackage{amsfonts}

\usepackage{lmodern} 
\usepackage{amsmath}                                                    
\usepackage{amsthm}                                                     
\usepackage{amssymb}
\usepackage{multirow}
\usepackage[colorlinks=true,citecolor=violet,urlcolor=violet,linkcolor=magenta]{hyperref}
\usepackage[nameinlink]{cleveref}
\Crefname{figure}{Fig.}{Figs.}

\usepackage{adjustbox}
\usepackage{color, colortbl} 
\definecolor{Gray}{gray}{0.9}
\usepackage{fancyhdr}
\usepackage{ragged2e}
\DeclareCaptionJustification{justified}{\justifying}
\usepackage{titlesec}
\begin{document}

	\title{\Large \bf  { $\mu$-hybrid Inflation, Gravitino Dark Matter and Stochastic Gravitational Wave Background from Cosmic Strings}}
\date{\today}
	
\author{\bf Adeela Afzal}
\affiliation{Department of Physics, Quaid-i-Azam University, Islamabad, 45320, Pakistan}
	\author{\bf Waqas Ahmed}
	\affiliation{School of Mathematics and Physics, Hubei Polytechnic University,	Huangshi 435003, China}
	\author{\bf Mansoor Ur Rehman}
	\email[E-mail: ]{mansoor@qau.edu.pk}
	\affiliation{Department of Physics, Quaid-i-Azam University, Islamabad, 45320, Pakistan}
	\author{\bf Qaisar Shafi}
	\affiliation{Bartol Research Institute, Department of Physics and Astronomy, University of Delaware, Newark, DE 19716, USA}
\begin{abstract}
	
	\noindent
	We present a successful realization of supersymmetric $\mu$-hybrid inflation model based on a gauged $U(1)_{B-L}$ extension of the minimal supersymmetric standard model, with the soft supersymmetry breaking terms are playing an important role. Successful non-thermal leptogenesis with gravitino dark matter yields a reheat temperature in the range $2 \times 10^{7} \lesssim T_R \lesssim 5 \times 10^{9}$~GeV. This corresponds to the predictions $2 \times 10^{-18} \lesssim r\lesssim 4 \times 10^{-13}$ for the tensor to scalar ratio, and $-2 \times 10^{-6} \lesssim dn_s/d\ln k \lesssim -5 \times 10^{-11}$ for the running of the scalar spectral index. The $B-L$ breaking scale is estimated as $ 6 \times 10^{14}\lesssim M/ \text{GeV}\lesssim 10^{16}$, calculated at the central value of the scalar spectral index, $n_s =0.9655$, reported by Planck 2018. Finally, in a grand unified theory setup the dimensionless string tension parameter associated with the metastable strings is  in the range $ 10^{-9} \lesssim G\mu_\text{cs} \lesssim 10^{-6}$ corresponding to a stochastic gravitational wave background lying within the 2$\sigma$ bound of the recent NANOGrav 12.5-yr data.
\end{abstract}
\maketitle

\section{\large{\bf Introduction}}
Supersymmetric (SUSY) hybrid inflation \cite{Dvali:1994ms,Copeland:1994vg,Linde:1997sj,Senoguz:2004vu,Buchmuller:2000zm,Senoguz:2003zw,Rehman:2009nq,Rehman:2009yj} offers an attractive framework for linking inflation with particle physics models based on grand unified theories (GUTs)  \cite{Senoguz:2003zw}. In the minimal models \cite{Rehman:2009nq,Rehman:2009yj}, the soft supersymmetry breaking terms play an important role in making the predictions for the scalar spectral index $n_s$ consistent with the cosmic microwave background (CMB) data \cite{Planck:2018jri,Planck:2018vyg}. Alternatively, the non-minimal terms in the K\"ahler potential serve a similar purpose \cite{Bastero-Gil:2006zpr,urRehman:2006hu,Shafi:2010jr,Rehman:2010wm,Ahmed:2022vlc}. With regard to linking inflation with particle physics models, $\mu$-hybrid inflation \cite{Dvali:1997uq,Okada:2015vka} offers an interesting class of SUSY hybrid inflation models where the $\mu$-problem of the minimal supersymmetric standard model (MSSM) is also resolved \cite{Dvali:1997uq,King:1997ia}. It is shown in \cite{Okada:2015vka} that $\mu$-hybrid inflation based on a renormalizable superpotential and minimal (canonical) K\"ahler potential leads to split supersymmetry scale with the gravitino mass, $m_{3/2} \lesssim 5 \times 10^{7}$~GeV. However, with non-minimal K\"ahler potential we can realize $\mu$-hybrid inflation with $m_{3/2} \sim 1-100$~TeV and reheat temperature $T_R \gtrsim 10^{6}$~GeV \cite{Rehman:2017gkm}. A shifted version of $\mu$-hybrid inflation is investigated in \cite{Lazarides:2020zof} where the monopole problem associated with the breaking of the underlying gauge symmetry can be avoided.  A viable model for gravitino dark matter and potentially detectable primordial gravitational waves are among the attractive features of these models. A discussion of successful leptogenesis in $\mu$-hybrid inflation,  however, is missing in these papers which is one of the motivations of this paper.

A minimal version of $\mu$-hybrid inflation is considered in the present paper with renormalizable superpotential and minimal (canonical) K\"ahler potential. As compared to an earlier treatment of this model in \cite{Rehman:2017gkm}, we here allow the soft SUSY breaking mass, $M_S$, to be different from the gravitino mass $m_{3/2}$, as assumed in \cite{Rehman:2009yj} for standard SUSY hybrid inflation model. This leads to interesting consequences related to the viability of $\mu$-hybrid inflation with gravitino dark matter and successful non-thermal leptogenesis. An adequate range of reheat temperature is obtained while avoiding the gravitino overproduction problem \cite{ELLIS1984181,Khlopov:1984pf}. 
\par
This realization of $\mu$-hybrid inflation is based on a gauged $U(1)_{B-L}$ extension of MSSM, where $B$ and $L$ are the baryon and lepton numbers respectively.
The breaking of $U(1)_{B-L}$ gives rise to a topologically stable cosmic string network that is usually constrained from the various experimental bounds. However, here we consider metastable cosmic strings where these bounds can be relaxed. A brief discussion related to the formation of such a metastable string network is presented in a GUT setup based on $SO(10)$.  This metastable cosmic string network can decay via the Schwinger production of monopole-antimonopole pairs \cite{Buchmuller:2019gfy} while generating a stochastic gravitational wave background (SGWB) in a range accessible at the ongoing and future gravitational wave (GW) experiments. We compare our model predictions with the recent bounds from the North American Nanohertz Observatory for Gravitational Waves (NANOGrav) 12.5-yr data \cite{NANOGrav:2020bcs}. In addition, we highlight the parameter space which is also consistent with gravitino dark matter and successful leptogenesis. For a similar study in no scale inflation see \cite{Ahmed:2022rwy}.

\section{\large{\bf Supersymmetric $\mu$-hybrid Inflation}}
The superpotential for hybrid inflation in a $U(1)_{B-L}$ extension of minimal supersymmetric standard model (MSSM) can be written as \cite{Senoguz:2005bc,Pallis:2017xfo,Masoud:2021prr},
\begin{align} \label{superpot1}
W&=\kappa S(\Phi\bar{\Phi}-M^2)+\lambda S H_u H_d
+y_{ij}^{(U)}Q_i U^c_j H_u  \notag \\
&+ y_{ij}^{(D)}Q_i D^c_j H_d+y_{ij}^{(L)} L_i E^c_j H_d 
+ y_{ij}^{(\nu)} L_i H_u N_j^c \notag \\ 
&+ \frac{\lambda_{ij}}{M_*} \bar{\Phi}^2 N^c_i N^c_j\,,  
\end{align}
where, $i,j=1,2,3$ are the family indices,
$\kappa,\lambda$ and $\lambda_{ij}$ are dimensionless couplings and $y_{ij}^{(U)},y_{ij}^{(D)},y_{ij}^{(L)}$ and $y_{ij}^{(\nu)}$ are the Yukawa couplings, 
involving MSSM superfields, ($Q_i$, $U^c_i$, $D^c_i$, $L_i$, $E^c_i$), with right-handed neutrino superfield, $N^c_i$, and the electroweak Higgs doublet superfields, ($H_u,\,H_d$). The last term, relevant for the right-handed neutrino masses, contains a cutoff scale, $M_*$.  
The scalar component of the gauge singlet superfield $S$ acts as an inflaton, and the $B-L$ conjugate pair of Higgs superfields ($\Phi$, $\overline{\Phi }$) provides the vacuum energy, $\kappa^2 M^4$, for inflation containing the $B-L$ symmetry breaking scale $M$. The above superpotential not only possesses the local $U(1)_{B-L}$ symmetry, but it also contains three global symmetries, namely, $U(1)_R$, $U(1)_B$ and $U(1)_L$, with $R(W)=2$ and $B(W)=0=L(W)$.  The charge assignments under these symmetries of the various matter and Higgs superfields are given in \cref{chargeassign}. 
\begin{table}[b]
	
	\caption{Global and local charges of superfields} 
	\centering 
	\begin{adjustbox}{max width=\columnwidth}
		\begin{tabular}{p{2cm} p{1.3cm} p{1.3cm} p{1.3cm} p{1.3cm}}
			\hline
			\bf{Superfields}& \bf{R}&\bf{B}&\bf{L}&\bf{B-L}
			\\ [0.5ex] 
			\hline\hline
			$E^c_i$ & 1 & 0 & -1 & 1\\ 
			$N^c_i$ & 1 & 0 & -1 & 1\\ 
			$L_i$ & 1 & 0 & 1 & -1\\ 
			$U^c_i$ & 1 & -1/3 & 0 & -1/3\\ 
			$D^c_i$ & 1 & -1/3 & 0 & -1/3\\ 
			$Q_i$ & 1 & 1/3 & 0 & 1/3\\ 
			$H_u$ & 0 & 0 & 0 & 0\\ 
			$H_d$ & 0 & 0 & 0 & 0\\ 
			$S$ & 2 & 0 & 0 & 0\\ 
			$\Phi$ & 0 & 0 & -1 & 1\\
			$\bar{\Phi}$ & 0 & 0 & 1 & -1\\
			\hline 
		\end{tabular}
	\end{adjustbox}
	\label{chargeassign}
\end{table}
\par

The global SUSY  minimum occurs at,
\begin{equation}
\langle S\rangle=0,  \,\,\,\,\,\, \langle\Phi\bar{\Phi}\rangle=M^2, \,\,\,\,\,\, \langle H_u\rangle = 0 = \langle H_d\rangle.
\end{equation}
After the breaking of $U(1)_{B-L}$ gauge symmetry, the last term in \cref{superpot1} gives rise to Majorana mass terms for the right-handed neutrinos,
\begin{equation}
M^R_{ij}=\lambda_{ij} \left( \frac{M}{M_*} \right)M.
\end{equation}
With $\lambda_{ij} \lesssim \mathcal{O}(1)$ and $M_* \sim 10^{18}$~GeV, we obtain Majorana masses $\lesssim 10^{14}$~GeV with the gauge symmetry breaking scale $M\sim10^{16}$~GeV. Therefore, the light neutrino masses are naturally generated via the seesaw mechanism. It is also interesting to note that the $R-$parity which prevents rapid proton decay mediated by the dimension four operators appears as a $Z_2$ subgroup of $U(1)_R$ symmetry. However, the proton is essentially stable due to the global and local symmetries
described in \cref{chargeassign}. A $Z_4$ subgroup of $U(1)_R$ symmetry, consistant with the $R$ charge assignment displayed in \cref{chargeassign}, is identified as a unique anomaly free discrete symmetry described in \cite{Lee:2010gv} that forbids both the $\mu$-term and all dimension four and five baryon and lepton number violating operators in MSSM.
\par
\begin{figure}[t]
	\centering
	\includegraphics[width=1.2\columnwidth]{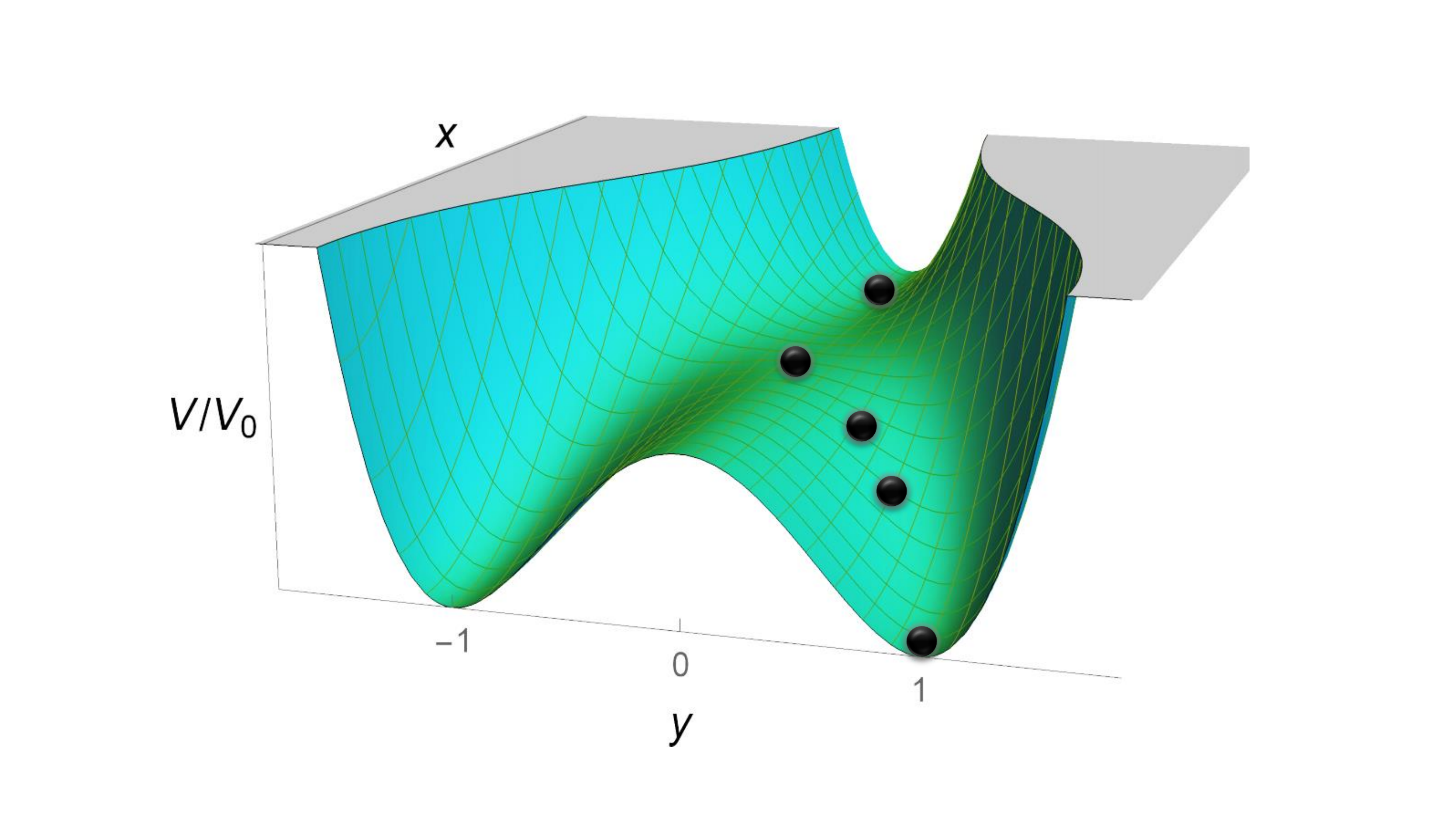}
	\caption{The normalized global SUSY scalar potential $V/V_0$ as a function of $x = |s|/M$ and $y = |\phi|/M$.}  \label{3Dplot}
	\label{fig:superpotential}
\end{figure}

The superpotential term relevant for hybrid inflation is \cite{Copeland:1994vg,Dvali:1994ms},
\begin{equation}
W = \kappa S(\Phi \overline{\Phi }-M^{2}),
\label{superpot}
\end{equation} 
and the global SUSY F-term scalar potential is given by,
\begin{equation}\label{VF1}
V_{F} = \kappa^2\,\vert M^2 - \phi\,\overline{\phi} \vert^2 + \kappa^2 \vert s \vert^2 (\vert \phi \vert^2+\vert \overline{\phi} \vert^2),
\end{equation}
where $\phi , \overline{\phi }, s$  represents the bosonic components of the superfields $\Phi, \overline{\Phi },S$ respectively. 
In the D-flat direction, $|\phi|=|\overline{\phi}|$, using \cref{superpot} and \cref{VF1}, we write the tree level global SUSY potential as,
\begin{equation}\label{eq1}
V = V_0 \left[ (1 - y^2)^2 + 2\, x^2 y^2 \right],
\end{equation}
where $V_0=\kappa^2 M^4$, $x = |s|/M$ and $y=|\phi|/M$. This scalar potential in displayed in \cref{3Dplot} where a flat direction ($y = 0$) with $x > 1$, suitable for inflation, is clearly visible. As described below, the various important contributions to the scalar potential provide the necessary slope for the realization of inflation in the otherwise flat trajectory. 

The F-term supergravity (SUGRA) scalar potential is given by,
\begin{equation}
V_{F}=e^{K/m_{P}^{2}}\left(
K_{ij}^{-1}D_{z_{i}}WD_{z^{*}_j}W^{*}-3m_{P}^{-2}\left| W\right| ^{2}\right),
\label{VF}
\end{equation}
where, 
\begin{equation*}
D_{z_{i}}W \equiv \frac{\partial W}{\partial z_{i}}+m_{P}^{-2}\frac{\partial K}{\partial z_{i}}W, \,\,\,\,\,\, K_{ij} \equiv \frac{\partial ^{2}K}{\partial z_{i}\partial z_{j}^{*}},
\end{equation*}
$D_{z_{i}^{*}}W^{*}=\left( D_{z_{i}}W\right) ^{*}$, $z_i \in \{\phi , \overline{\phi }, s \}$, $z_i^*$ being conjugate, and $m_{P} \simeq 2.4\times 10^{18}$~GeV is the reduced Planck mass. In the present paper we employ the minimal (canonical) K\"ahler potential given by,
\begin{equation}
K=  |S|^{2}+ |\Phi|^{2} + |\overline{\Phi}|^{2},  \label{kahler}
\end{equation}
and the SUGRA corrections can now be calculated using the above definitions. 
Along the inflationary trajectory SUSY is broken due to the non-zero vacuum term, $V_0$. This generates a mass splitting between the fermionic and the bosonic  components of the relevant superfields and leads to radiative corrections in the scalar potential \cite{Dvali:1994ms}. Another important contribution in the scalar potential arises from the soft SUSY breaking terms \cite{Senoguz:2004vu,Buchmuller:2000zm,Rehman:2009nq}.
\par 
Including the leading order SUGRA corrections, one-loop radiative corrections and the soft SUSY breaking terms, the scalar potential along the inflationary trajectory (i.e. $y=0$) can be written as \cite{Rehman:2009nq,urRehman:2006hu,Rehman:2010wm},
\begin{align}\label{scalpot}
V  &\simeq
V_0 \left[ 1 + \left( \frac{M}{m_{P}}\right) ^{4}\frac{x^{4}}{2}+\frac{\kappa ^{2}}{8\pi ^{2}}F(x) + \frac{\lambda ^{2}}{4\pi ^{2}}F(\sqrt{\gamma}\,x) \right. \notag \\
&\left.+ a\left(\frac{m_{3/2}\,x}{\kappa\,M}\right) + \left( \frac{M_S\,x}{\kappa\,M}\right)^2 \right],
\end{align}
where $\gamma \equiv \lambda/\kappa$ and
\begin{align}
F(x)&=\frac{1}{4}\left[ \left( x^{4}+1\right) \ln \frac{\left( x^{4}-1\right)}{x^{4}}+2x^{2}\ln \frac{x^{2}+1}{x^{2}-1}\right. \notag\\
&\left.+2\ln \frac{\kappa ^{2}M^{2}x^{2}}{Q^{2}}-3\right],
\end{align}
is the one-loop radiative correction function evaluated at the renormalization scale $Q$, and $a$ is defined as,
\begin{equation}
a = 2\left| 2-A\right| \cos [\arg s+\arg (2-A)].
\label{a}
\end{equation}
\par
The last two terms in \cref{scalpot} are the soft SUSY breaking linear and mass-squared terms, respectively, obtained in a gravity-mediated SUSY breaking scheme.
The presence of $a$-term makes the present model a two-field inflation model \cite{Buchmuller:2014epa}. However, we assume a suitable initial condition for $\arg s$ so that $a$ remains constant during inflation \cite{urRehman:2006hu}. We further assume the soft mass, $M_S$, of the $s$ field to be different, in general, from the gravitino mass, $m_{3/2}$. As we will see later, this choice provides an extra degree of freedom which yields a relatively wider range of $M$ consistent with the central value of the spectral index $n_s = 0.966$ measured by  Planck 2018 \cite{Planck:2018vyg}. 
The dimensionless parameter $a$ and the soft mass squared, $M_S^2$, can have any sign. For standard SUSY hybrid inflation, it is shown in \cite{Rehman:2009nq}  that choosing the negative sign for either soft SUSY breaking term predicts the scalar spectral index $n_s$ in good agreement with the central value reported by Planck 2018. We envisage similar results in the present $\mu$-hybrid inflation model.
\section{\large{\bf Inflationary Observables in Slow-roll Approximation}}
The prediction for the various inflationary parameters are estimated using the standard slow roll parameters defined as,
\begin{align}
\epsilon &= \frac{1}{4}\left( \frac{m_P}{M}\right)^2
\left( \frac{V'}{V}\right)^2, \,\,\,
\eta = \frac{1}{2}\left( \frac{m_P}{M}\right)^2
\left( \frac{V''}{V} \right), \notag\\
\xi^2 &= \frac{1}{4}\left( \frac{m_P}{M}\right)^4
\left( \frac{V' V'''}{V^2}\right),
\label{slowroll}
\end{align}
where prime denotes the derivative with respect to $x$. Note that the extra factor of $1/2$ is due to the relation between the canonically normalized real inflaton field, $\sigma \equiv |s|/\sqrt{2}$, and the complex field, $s$. In the slow-roll approximation, the scalar spectral index $n_s$, the tensor-to-scalar ratio $r$ and the running of the scalar spectral index $\alpha_{s}\equiv dn_s / d \ln k$ are given by,
\begin{align} \label{nsr}
n_s&\simeq 1+2\,\eta-6\,\epsilon, \,\,\,\,\,\,\,\,\,\,\,
r \simeq 16\,\epsilon, \notag\\
\alpha_{s} &\simeq 16\,\epsilon\,\eta
-24\,\epsilon^2 - 2\,\xi^2.
\end{align}
The value of the scalar spectral index $n_s$ in the $\Lambda$CDM model is $n_s = 0.9665 \pm 0.0038$ \cite{Planck:2018jri}.

The amplitude of the scalar power spectrum is given by,
\begin{align}
A_{s}(k_0) = \frac{1}{24\,\pi^2\,\epsilon(x_0)}
\left( \frac{V(x_0)}{m_P^4}\right),  \label{curv}
\end{align}
where  $A_{s}(k_0) = 2.137 \times 10^{-9}$ at the pivot scale $k_0 = 0.05\, \rm{Mpc}^{-1}$ as measured by Planck 2018 \cite{Planck:2018jri}.
The relevant number of e-folds, $N_0$, before the end of inflation is,
\begin{align}\label{Ngen}
N_0 = 2\left( \frac{M}{m_P}\right) ^{2}\int_{x_e}^{x_{0}}\left( \frac{V}{%
	V'}\right) dx,
\end{align}
where $x_0 \equiv x(k_0)$ is the field value at the pivot scale $k_0$, and
$x_e$ is the field value at the end of inflation. 
As the case may be, the value of $x_e$ is fixed either by the breakdown of the slow roll approximation ($\eta(x_e)=-1$), or by a `waterfall' destabilization occurring at the value $x_e = 1$. 
\begin{figure*}[t!]
	\centering
	\subfloat[\label{sub:MS}]
	{{\includegraphics[width=0.921\columnwidth]{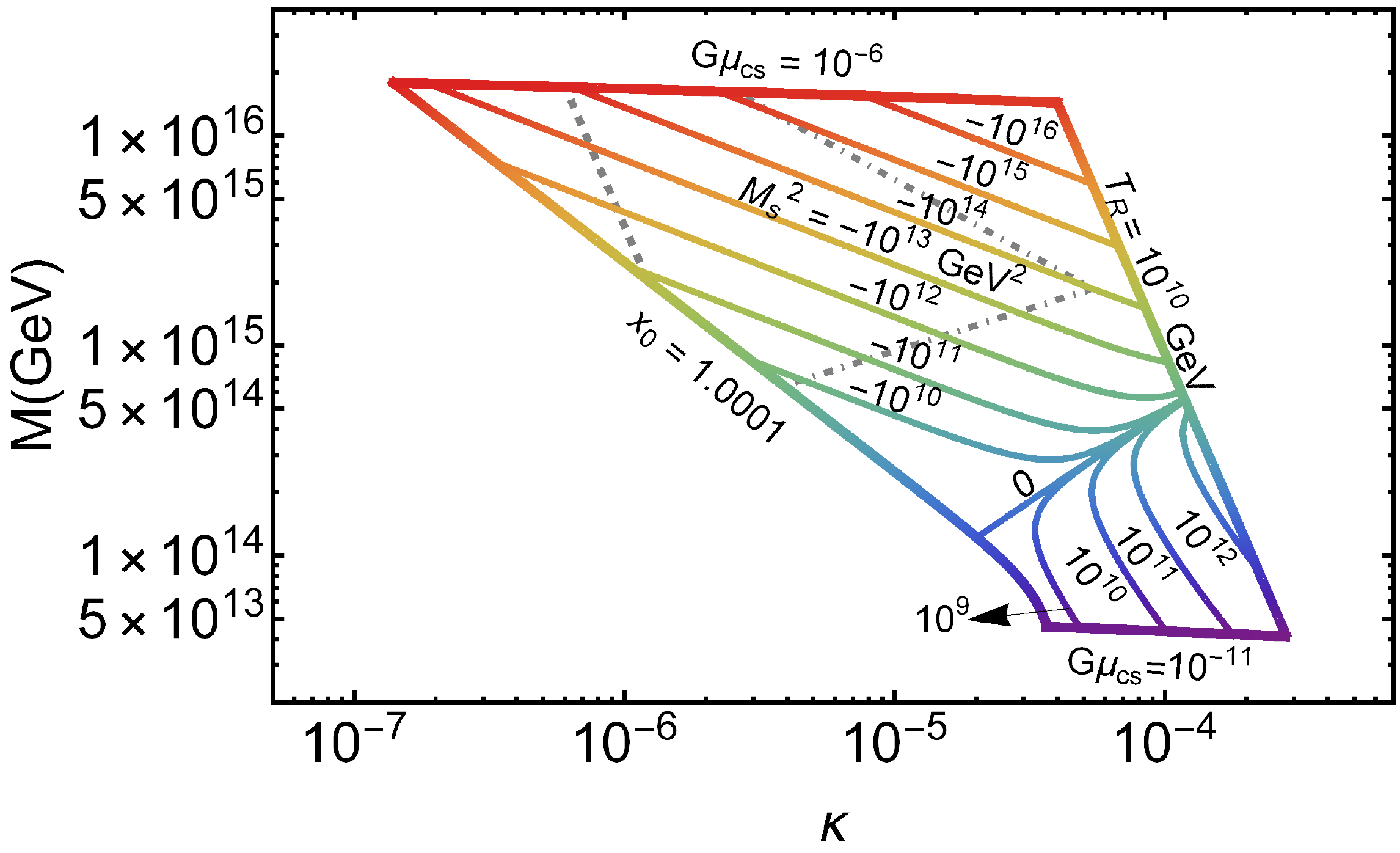} }}%
	\quad
	\subfloat[\label{sub:m32}]
	{{ 	\includegraphics[width=1.06\columnwidth]{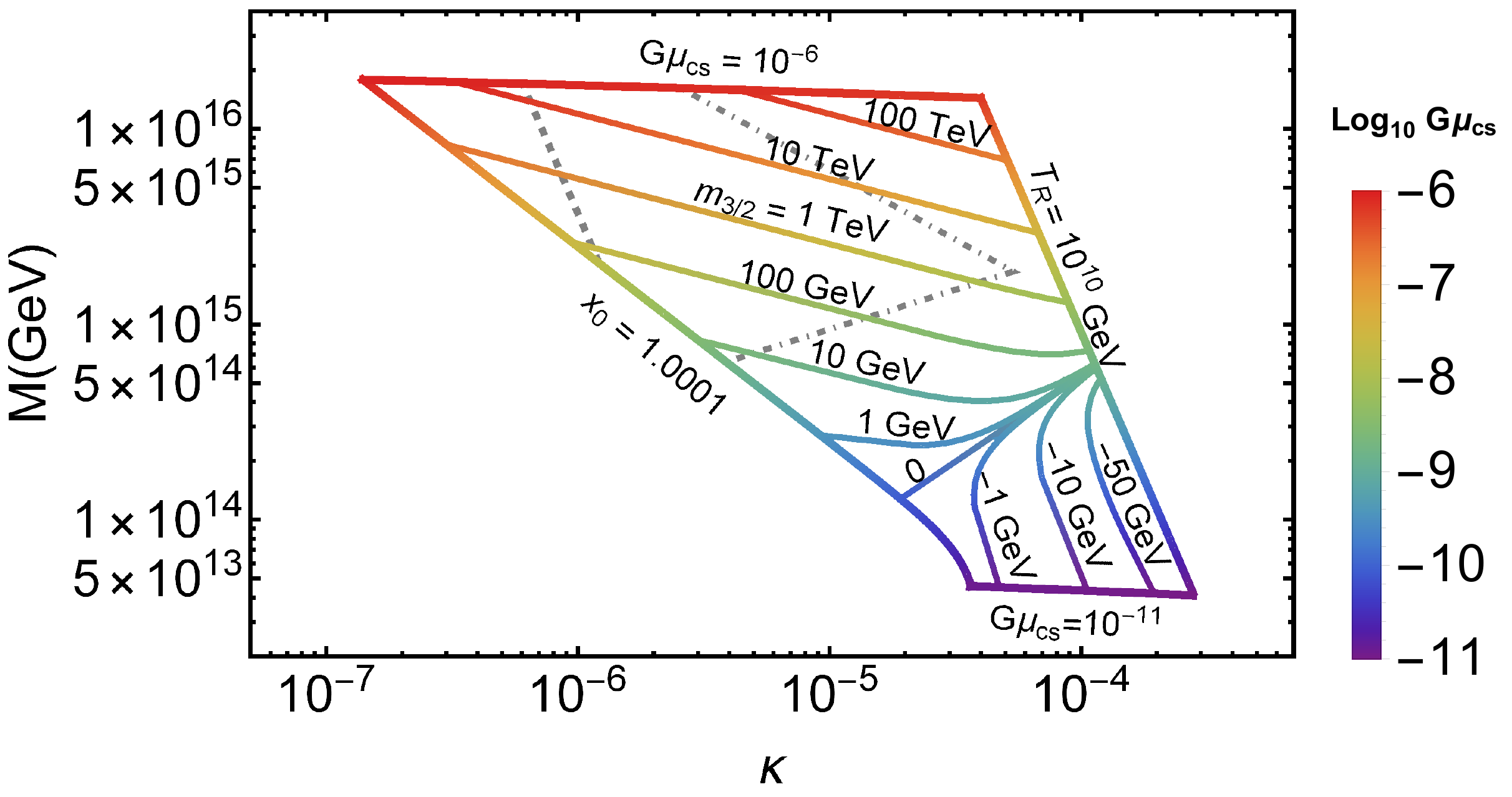} }}%
	\caption{The symmetry breaking scale $M$ versus the coupling $\kappa$, with maximum (minimum) reheat temperature $T_R$ of $10^{10}$ GeV ($2.3\times10^6$ GeV), and fine tuning bound of 0.01\%. The rainbow color vertical bar represents the variation of the dimensionless string tension $G\mu_\text{cs}$  from $10^{-6}$ to $10^{-11}$. The inside mesh shows the variation of soft mass term $M_S^2$ in \ref{sub:MS}, and that of gravitino mass $m_{3/2}$, in \ref{sub:m32}. Gray dashed line is the $T_R$ bound of leptogenesis, and dot-dashed is of stable LSP gravitino DM.}
	\label{fig:reheatr}
\end{figure*}
\section{\large{\bf Reheating and Leptogenesis}}
After the end of inflation, the system falls towards the SUSY vacuum and performs damped oscillations about it. The inflaton  consists of the two complex scalar fields $s$ and $\theta=\left(\delta  \phi + \delta \bar{\phi}\right)/\sqrt{2}$ with the same mass,
$m_\text{inf} \simeq \sqrt{2}\kappa M$.
The inflaton predominantly decays into a pair of higgsinos ($\widetilde h_u$, $\widetilde h_d$) and higgses ($ h_u$, $ h_d$), each with a decay width, $\Gamma_h$, given by \cite{Lazarides:1998qx},
\begin{equation}\label{gamma}
\Gamma_h =\Gamma(\theta \rightarrow \widetilde h_u\widetilde h_d) = \Gamma(s \rightarrow h_u h_d)=\frac{\lambda ^2 }{8 \pi }m_{\text{inf}}.
\end{equation}
The other decay mode, via the superpotential couplings $(\lambda_{ij}/M_*) \, \bar{\Phi}^2  N^{c}_i N^{c}_j$, leads to a pair of right-handed neutrinos ($N$) and sneutrinos ($\widetilde{N}$) respectively with equal decay width given as, 
\begin{align}
\Gamma_N& = \Gamma (\theta \rightarrow NN )= \Gamma(s \rightarrow \widetilde{N}\widetilde{N} ) \notag\\
&=\dfrac{m_\text{inf}}{8\pi}\left( \frac{M_N}{M} \right)^2\left(1-\dfrac{4M_{N}^{2}}{m_\text{inf}^{2}}\right)^{1/2},
\end{align}
provided that only the lightest right-handed neutrino with mass $M_{N}$ satisfies the kinematic bound, $m_\text{inf} > 2 M_{N}$. 

The relevant Boltzmann equations for the evolution of the total energy density, $\rho$, of $s$ and $\theta$ fields and the radiation energy density, $\rho_r$, are given by,
\begin{equation}
\dot{\rho} = -3 H \rho - \Gamma_\text{inf} \,\rho, \quad
\dot{\rho}_r = -4H \rho_r + \Gamma_\text{inf} \, \rho_r , 
\end{equation}
where,
\begin{equation}
H^2 = \dfrac{\rho + \rho_r}{3 m_p^2}\,\,\,\ \text{ and } \,\,\,\ \Gamma_\text{inf} = \Gamma_h + \Gamma_N.
\end{equation}
With $H = 3\Gamma_\text{inf}$, we define the reheat temperature $T_R$ in terms of the inflaton decay width $\Gamma_\text{inf}$,
\begin{equation}\label{tr}
T_R=\left(\dfrac{90}{\pi^{2}g_{*}}\right)^{1/4}\sqrt{\Gamma_\text{inf} \, m_{P}},
\end{equation}
where $g_{*}= 228.75$ for MSSM. Assuming a standard thermal history, the number of e-folds, $N_{0}$, can be written in terms of the reheat temperature, $T_R$, as \cite{Liddle:2003as},
\begin{align}\label{efolds}
N_0=53+\dfrac{1}{3}\ln\left[\dfrac{T_R}{10^9 \text{ GeV}}\right]+\dfrac{2}{3}\ln\left[\dfrac{\sqrt{\kappa}\,M}{10^{15}\text{ GeV}}\right].
\end{align}

Note that the effect of preheating in supersymmetric hybrid inflation is generally  expected to be suppressed \cite{Garcia-Bellido:1997hex, Bastero-Gil:1999sik}. However, if both the inflaton and waterfall fields  are coupled to an additional scalar field, the preheating can be efficient if the inflaton is relatively strongly coupled to this scalar field \cite{Garcia-Bellido:1997hex}. In the present case, the electroweak Higgs doublet in the D-flat direction represents such a scalar field and efficient preheating requires $\lambda >> \kappa$. As we only consider $\lambda \sim \kappa$, the non-perturbative effects via preheating are expected to be suppressed in our case. In addition, preheating associated with fermions is generically expected to be subdominant due to Pauli blocking.

\begin{figure*}[t]
	\centering
	\subfloat[\label{sub:Tr}]
	{{\includegraphics[width=0.92\columnwidth]{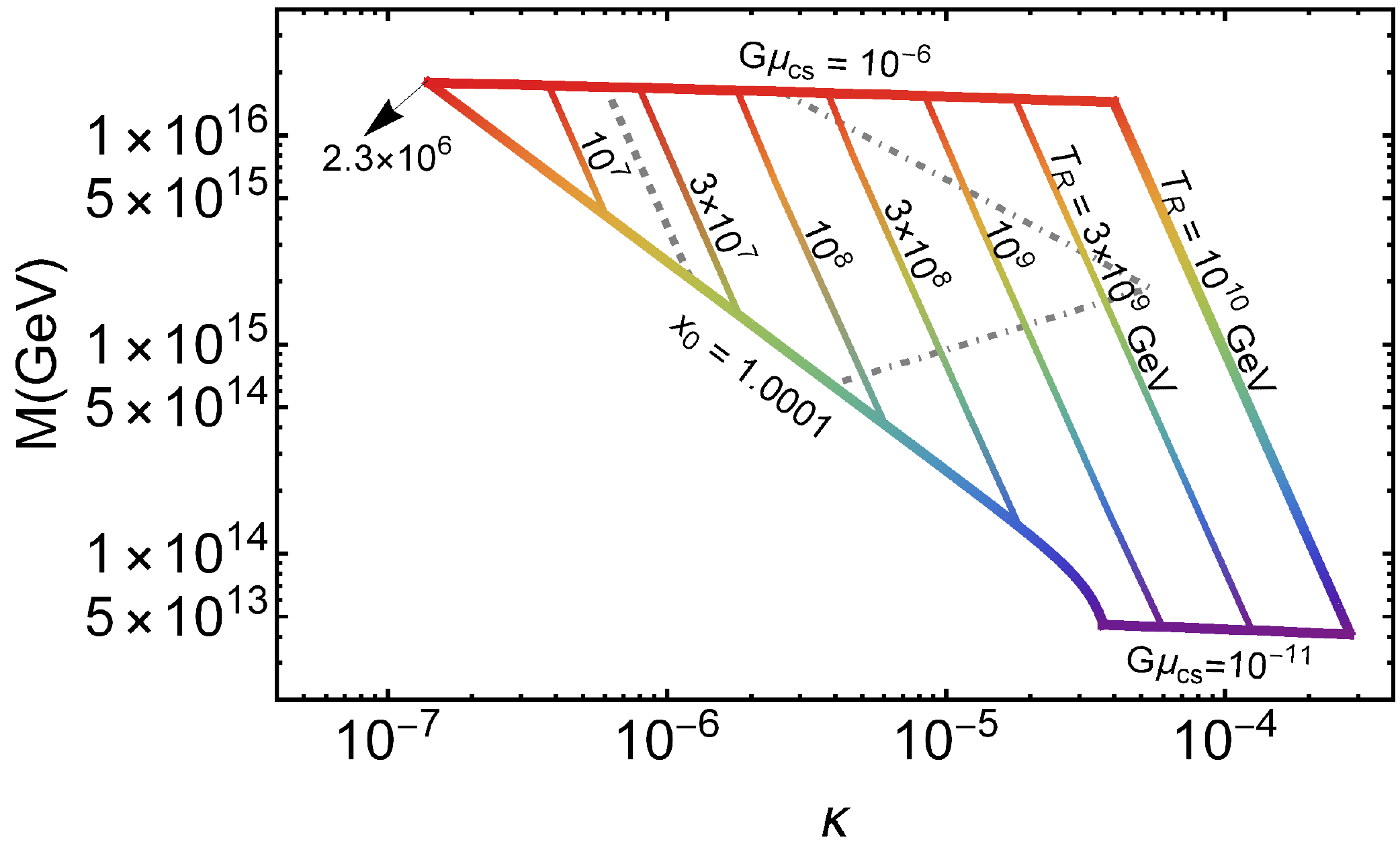} }}%
	\quad
	\subfloat[\label{sub:minf}]
	{{ 	\includegraphics[width=1.06\columnwidth]{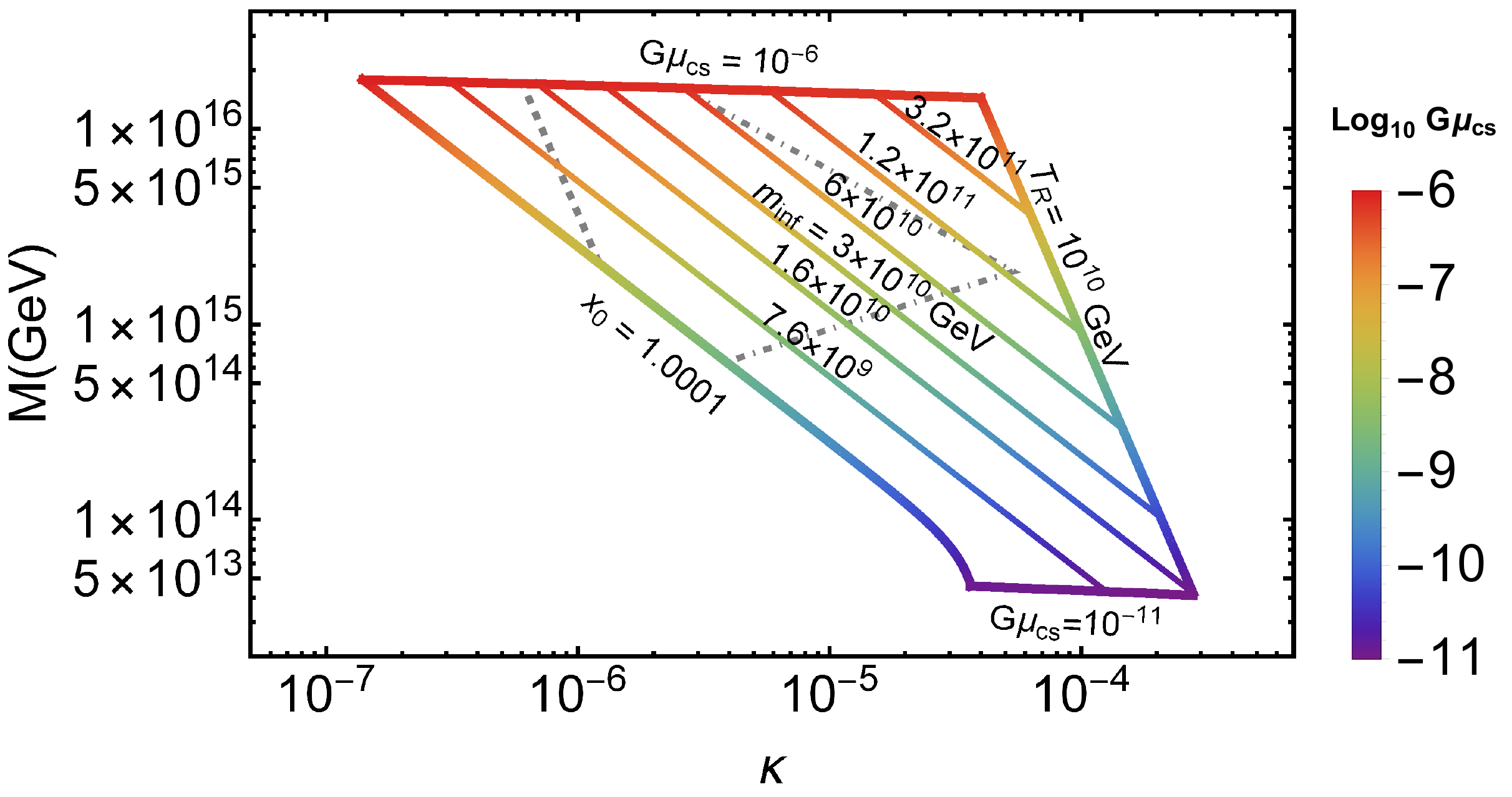} }}%
	\caption{The symmetry breaking scale $M$ versus the coupling $\kappa$, with maximum (minimum) reheat temperature $T_R$ of $10^{10}$ GeV ($2.3\times10^6$ GeV), and fine tuning bound of 0.01\%. The rainbow color vertical bar represents the variation of the dimensionless string tension $G\mu_\text{cs}$  from $10^{-6}$ to $10^{-11}$. The inside mesh shows the variation of $T_R$ in \ref{sub:Tr}, and that of inflaton mass $m_{\text{inf}}$ in \ref{sub:minf}. Gray dashed line is the $T_R$ bound of leptogenesis, and dot-dashed is of stable LSP gravitino DM.}
	\label{fig:reheatr2}
\end{figure*}

Although sub-dominant, $\Gamma_N /\Gamma_h \leq 1/(3\sqrt{3} \gamma^2) \simeq (0.4/\gamma)^2 < 1 $, the $\Gamma_N$ channel is important for successful leptogenesis  which is partially converted into the observed baryon asymmetry through the sphaleron process \cite{Kuzmin:1985mm,Fukugita:1986hr,Khlebnikov:1988sr}. The washout factor of lepton asymmetry can be suppressed by assuming  $M_N \gg T_R$. The observed baryon asymmetry is evaluated in term of the lepton asymmetry factor, $\varepsilon_L$,
\begin{align}\label{bphr}
\frac{n_{B}}{n_{\gamma}}\simeq -1.84 \,\varepsilon_L  \frac{\Gamma_N}{\Gamma_\text{inf}}\frac{T_R}{m_\text{inf}} \delta_{eff},
\end{align} 
where $\delta_{eff}$ is the CP violating phase factor, $\Gamma_\text{inf} \simeq \Gamma_h $ and, assuming hierarchical neutrino masses, $\varepsilon_L$ is given by \cite{Rehman:2018gnr},
\begin{eqnarray}
(-\varepsilon_L)  \simeq \frac{3}{8\pi}  \frac{\sqrt{\Delta m_{31}^{2}} M_N}{\langle H_{u}\rangle^{2}}.
\end{eqnarray} 
Here, the atmospheric neutrino mass squared difference is $\Delta m_{31}^{2}\approx 2.6 \times 10^{-3}$ eV$^{2} $  and $\langle H_{u}\rangle \simeq 174$ GeV in the large $\tan\beta$ limit. For the observed baryon-to-photon ratio,  $n_\mathrm{B} /n_\gamma = (6.12 \pm 0.04) \times 10^{-10}$ \cite{ParticleDataGroup:2020ssz}, the bound on $|\delta_{eff}|\leq 1$ along with the kinematic bound, $m_\text{inf} \geq 2M_N$, translates into a bound on the reheat temperature, 
\begin{align}\label{lept}
T_R  \gtrsim \gamma^2 \, 2 \times 10^{7} \text{ GeV} \geq 2 \times 10^{7} \text{ GeV},
\end{align} 
for $\gamma \geq 1$. In order to attain the minimum possible reheat temperature we set $\gamma=1$. This bound from successful leptogenesis, i.e $T_R\gtrsim 2 \times 10^7$~GeV, is represented by the gray dashed line in \cref{fig:reheatr,fig:reheatr2,fig:reheatr3}. An underproduction of  leptogenesis is assumed for $T_R < 2 \times 10^{7} \text{ GeV}$ with a reduction in $M_N$. 

\section{\large{\bf Cosmic String Constraints}}
Cosmic Strings (CSs) are produced at the end of inflation with implications related to anisotropies in the CMB and the production of stochastic gravitational waves (SGWs). The predictions related to SGWs are discussed in \cref{csdis}. The strength of the string's gravitational interaction is expressed in terms of the dimensionless string tension, $G\mu_\text{cs}$, where $G=1/8\pi m_P^2$ and $\mu_\text{cs}$ is mass per unit length of the string. The CMB bound on the CS tension is \cite{Planck:2018vyg,Planck:2018jri},
\begin{equation}
G\mu_\text{cs}\lesssim 1.3 \times 10^{-7}.
\end{equation}
The quantity $\mu_\text{cs}$, can be written in terms of the $U(1)_{{B-L}}$ gauge
symmetry breaking scale $M$ \cite{PhysRevLett.59.2493},
\begin{align}\label{mucs}
\mu_\text{cs}&= 2\pi M^2 \epsilon(\beta), \notag\\	\epsilon(\beta)&=\frac{2.4}{\log(2/\beta)} \text{  for  } \beta=\frac{\kappa^2}{2g^2}<10^{-2},
\end{align}
where $g=0.7$ for MSSM. Requiring $M \lesssim 10^{16}$ GeV, $G\mu_\text{cs} \lesssim 10^{-6}$ and this is possible with a metastable CS network as described in \cite{Buchmuller:2019gfy}. This possibility not only circumvents the CMB bound on $G\mu_\text{cs}$, it can also evade other bounds coming from LIGO O3 \cite{KAGRA:2021kbb}. A possible realization of a metastable CS network in a GUT setup based on $SO(10)$ model is described in \cref{csdis}.
\section{\large{\bf Results and Discussion}}
The seven parameters of the present model, $\kappa,\, M,\, am_{3/2},\, M_S^2,\, x_0,\, x_e,\, \text{and}\, M_N $, are constrained by five conditions, namely, the amplitude of scalar power spectrum, $A_s(k_0)=2.137\times 10^{-9}$ \cref{curv}, the scalar spectral index, $n_s = 0.9665$, the end of inflation by the waterfall, $x_e=1$, the number of e-folds, $N_0$, defined in \cref{Ngen} and given in terms of $T_R$ by \cref{efolds}, and finally the observed value of baryon-to-photon ratio, $n_B/n_{\gamma}=6.12\times 10^{-10}$ \cref{bphr}. This leaves two independent parameters to freely vary which can be taken to be $am_{3/2}$ and $M_S^2$. Keeping one parameter fixed, the second can be varied as depicted in \cref{sub:MS,sub:m32}. 
In \cref{sub:MS}, we vary $am_{3/2}$ for various values of $M_S^2$ in the range $2\times10^{12}\, \text{GeV}^2$ to $-2\times10^{17}\, \text{GeV}^2$. In \cref{sub:MS}, the curve with $M_S \sim 0$ has already been discussed in \cite{Rehman:2017gkm} with a minimal (canonical) K\"ahler potential. On the other hand, the region with $|M_S^2| \neq 0$ is the new parametric space explored in this paper. However, see \cite{Rehman:2009yj} where this region is explored in the standard hybrid inflation model ($\gamma \ll 1$). Similarly, in \cref{sub:m32} we vary $M_S^2$ for fixed values of $m_{3/2}$ lying in the range from $0$  to $730\, \text{TeV}$ ($0$ to $155$~GeV) for $a=1$ ($-1$). 

In accordance with the outcome of SUSY hybrid inflation model \cite{Rehman:2009nq,Rehman:2009yj}, at least one of the two parameters, $M_S^2$ or $am_{3/2}$, is expected to be negative in order to realize the red-tilted scalar spectral index consistent with Planck-2018 data. In the present model, the scalar spectral index with $x_0 \sim 1$ can be written as,
\begin{equation}\label{ns}
n_s \simeq 1+ \left(\dfrac{m_P}{M}\right)^2 \left( 2\left(\dfrac{M_S}{\kappa M}\right)^2 +3\dfrac{\kappa^2 }{8\pi^2}F''(x_0)\right).
\end{equation}
In the limit where $M_S^2$ term is dominant in the above expression we obtain,
\begin{equation}
  \left(\dfrac{M_S}{\kappa M}\right)^2 \simeq -\frac{(1-n_s)}{2} \left(\dfrac{M}{m_P}\right)^2.
\end{equation}
This explains the $M \propto \kappa^{-1/2}$ behavior of the curves in \cref{sub:MS} for most of the upper region with constant values of $M_S^2$. This behavior changes near the  $M_S \sim 0$ curve where the radiative term in \cref{ns} becomes dominant and thus predicting the $M \propto \kappa$ behavior. Regarding the behavior of the corresponding curves in \cref{sub:m32} with fixed values of $am_{3/2}>0$, it is noted that the soft SUSY breaking terms compete with each other in $\epsilon(x_0)$ in order to satisfy the constraint of $A_s$ given in \cref{curv}. Using this observation and \cref{ns} we obtain $M \propto \kappa^{-1/3}$ which is consistent with the curves shown in the upper region of \cref{sub:m32}.

\par 
For the case $M_S^2>0$, the radiative correction term starts to compete with the $M_S^2$ term in \cref{ns} while moving away from the $M_S \sim 0$ curve. This gives rise to $M \propto \kappa^{-2}$ behavior as can be seen in the lower region of \cref{sub:MS}.
Regarding the behavior of the corresponding region in \cref{sub:m32} with fixed values of $am_{3/2}>0$, it is noted that the soft SUSY breaking and radiative correction terms are comparable in $\epsilon(x_0)$ in order to satisfy the constraint of $A_s$ given in \cref{curv}. This leads to $M \propto \kappa^{-3}$ behavior of curves in the lower region of \cref{sub:m32}.

\begin{figure*}[t!]
	\centering
	\subfloat[\label{sub:r}]
{{ 	\includegraphics[width=0.91\columnwidth]{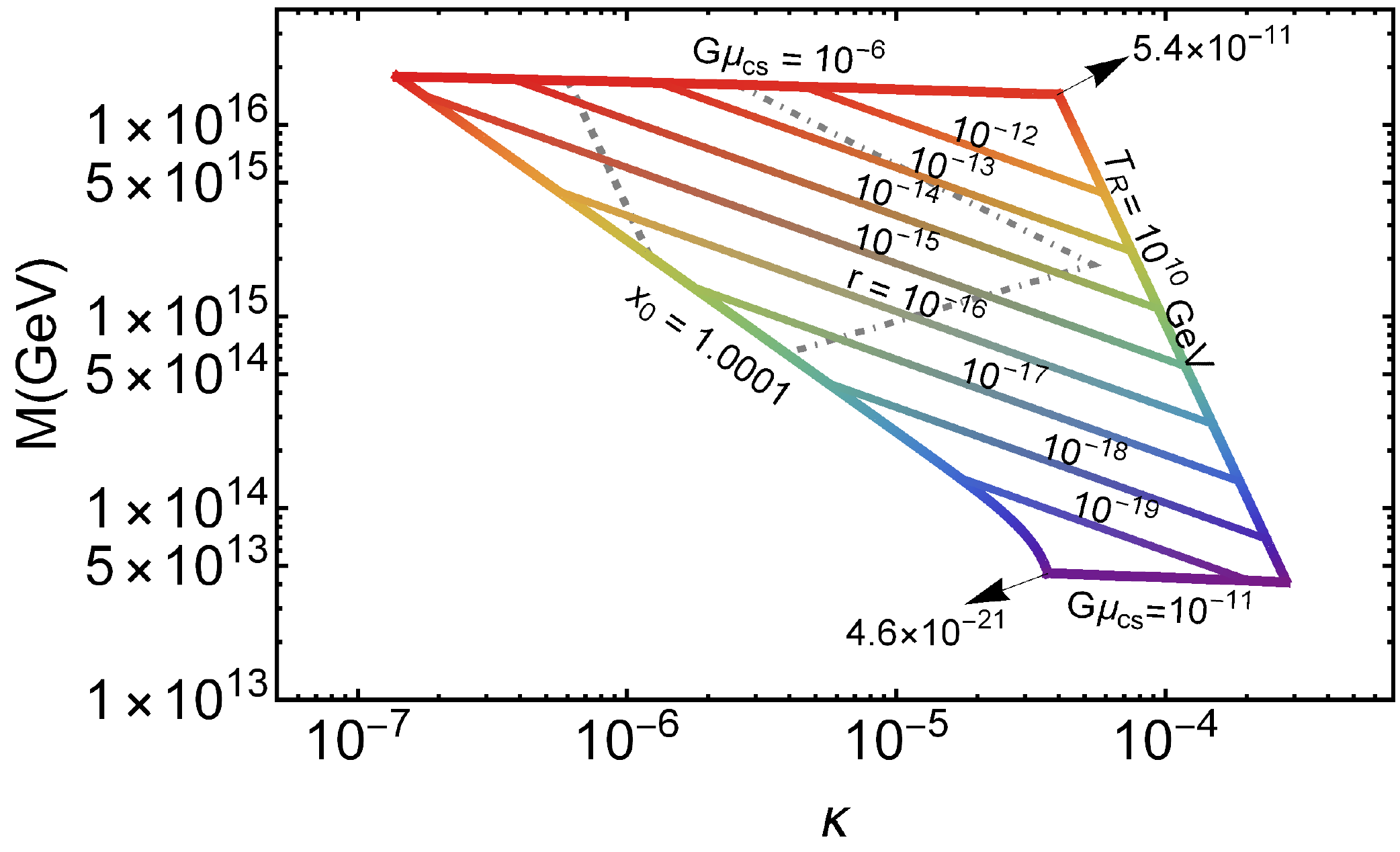} }}%
	\quad
\subfloat[\label{sub:alfa}]
{{ 	\includegraphics[width=1.04\columnwidth]{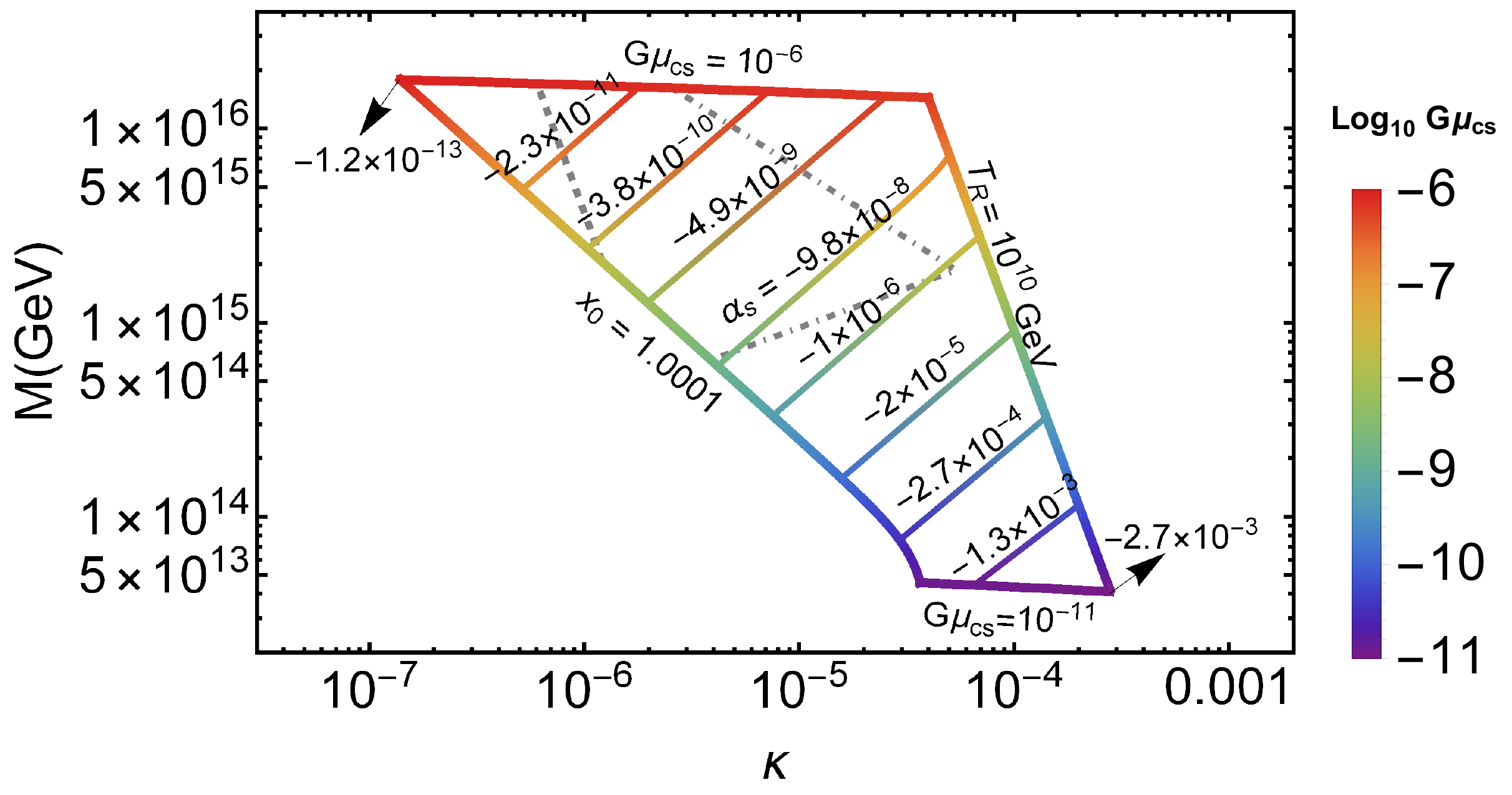} }}%
	\caption{The symmetry breaking scale $M$ versus the coupling $\kappa$, with maximum (minimum) reheat temperature $T_R$ of $10^{10}$ GeV ($2.3\times10^6$ GeV), and fine tuning bound of 0.01\%. The rainbow color vertical bar represents the variation of the dimensionless string tension $G\mu_\text{cs}$  from $10^{-6}$ to $10^{-11}$. The inside mesh shows the variation of tensor to scalar ratio $r$ in \ref{sub:r}, and that of the running of spectral index $\alpha_{s}$ in \ref{sub:alfa}. Gray dashed line is the $T_R$ bound of leptogenesis, and dot-dashed is of stable LSP gravitino DM.}
	\label{fig:reheatr3}
\end{figure*}

The four boundary curves in \cref{fig:reheatr,fig:reheatr2,fig:reheatr3} are respectively described by $G\mu_\text{cs} = 10^{-6}, \,10^{-11}$, $T_R=10^{10}$~GeV and $x_0=1.0001$. We do not consider larger values of reheat temperature, $T_R \gtrsim 10^{10}$~GeV, which are usually constrained by the gravitino overproduction problem and allow up to 0.01\% difference between $x_0$ and $x_e=1$, since for smaller values of $\kappa$ the corresponding field value $x_0$ happens to lie closer to the waterfall point, $x=1$. Owing to its direct dependence on $M$, the various fixed values of $G\mu$ are almost horizontal while having a weak dependence on $\kappa$. A wider range of the gauge symmetry breaking scale, $10^{13} \lesssim M/\text{GeV} \lesssim 10^{16}$ is obtained, as compared to $M_S \sim 0$ curve, where the range $(1-6)\, \times 10^{15}$~GeV is realized.
As shown in \cref{sub:Tr}, the curves with fixed values of reheat temperature, $T_R$, ranging between $2 \times10^6$~GeV to $10^{10}$~GeV follow $M \propto \kappa^{-3}$ behavior obtained from \cref{tr,gamma}. Further, the curves with fixed values of inflaton mass, $m_{\text{inf}} \simeq \sqrt{2} \kappa M$, ranging from  $2 \times10^9$~GeV to $8\times 10^{11}$~GeV are shown in \cref{sub:minf} and are consistent with $M \propto \kappa^{-1}$ behavior.
\par
The predicted range of the tensor to scalar ratio with tiny values, $r \sim 5\times10^{-11}-5\times10^{-21}$, is shown in \cref{sub:r} where the various curves with constant values of $r$ follow $M \propto \kappa^{-1/2}$ behavior as can be obtained from \cref{curv}, 
\begin{equation}
r \sim \frac{2}{3\,\pi^2 A_s(k_0)}\frac{\kappa^2 M^4}{m_P^4}.
\end{equation}
Finally, the relevant expression of $\alpha_s$ in the slow-roll approximation is given by,
\begin{align} \label{as}
\alpha_s& \simeq -\dfrac{1}{8}\left(\dfrac{m_P}{M}\right)^3\left(12 \left(\dfrac{ M}{m_P}\right)^4 x_0 +3\dfrac{\kappa^2}{8\pi^2}F'''(x_0)\right)\sqrt{r}.
\end{align}
The predicted range $-\alpha_s \sim 2.7\times10^{-3}-1.2\times10^{-13}$ is shown in \cref{sub:alfa} where the curves with constant values of $\alpha_s$ follow $M\propto \kappa$, that can be obtained from \cref{as} assuming a dominant contribution from the radiative correction with $F'''(x_0) \propto \kappa^{-2}$. The predicted ranges of $r$ and $\alpha_s$ with tiny values are consistent with the underlying assumption of the $\Lambda$CDM model.
\section{\large{\bf Gravitino Dark Matter}}
Following \cite{Rehman:2017gkm,Okada:2015vka,Lazarides:2020zof,Ahmed:2021dvo}, an interesting realization of  a stable gravitino as a cold dark matter (DM) candidate is presented here.
The relic abundance for stable gravitinos is described in terms of the reheat temperature $T_R$ and the gluino mass, $m_{\tilde{g}}$, as \cite{BOLZ2008336,Pradler:2006qh,Addazi:2017kbx,Eberl:2020fml},
\begin{align}\label{dmab}
\Omega_{3/2}h^2=0.08\left(\dfrac{T_R}{10^{10}~\text{GeV}}\right)\left(\dfrac{m_{3/2}}{1~ \text{TeV}}\right)\left(1+\dfrac{m_{\tilde{g}}^2}{3m^2_{3/2}}\right).
\end{align}
  This expression contains only the dominant QCD contributions for the gravitino production rate. The electroweak contribution \cite{Pradler:2006qh} is expected to be relatively suppressed for our analysis.
The observed DM abundance requires, $\Omega_{3/2}h^2\sim 0.12$ \cite{Planck:2018jri}, which allows us to write the gravitino mass in terms of the reheat temperature for a given value of gluino mass. The gray dot-dashed curves in \cref{fig:reheatr,fig:reheatr2,fig:reheatr3} are derived from  \cref{dmab} by taking into account $m_{\tilde{g}}\gtrsim m_{3/2}$ and the LHC bound on the gluino mass, $m_{\tilde{g}}> 2.2$~TeV \cite{Vami:2019slp}. The region to the left of these curves describes the gravitino DM in totality. It covers the values of gravitino mass in the range, $m_{3/2}\sim 6\,\text{GeV}-63\,\text{TeV}$ with reheat temperature up to $6\times10^9\,\text{GeV}$. Hence, the viable parameter space compatible with both DM and leptogenesis lies in the region bounded by the gray dashed and dot-dashed curves  of \cref{fig:reheatr,fig:reheatr2,fig:reheatr3}. Assuming an underproduction of leptogenesis, the region left of the gray dot-dashed curve (see \cref{fig:reheatr,fig:reheatr2,fig:reheatr3}) is also compatible with gravitino DM. 

With LSP gravitino the next to lightest supersymmetric particle (NLSP) $\widetilde{X}$ can decay into SM particles and gravitino. In this case the lifetime of $\widetilde{X}$ should be less than $1$~sec in order to keep the successful predictions of big bang nucleosynthesis (BBN) intact. The decay rate for $\widetilde{X}\longrightarrow\psi_\mu \gamma$ is given by \cite{Kawasaki:2008qe},
\begin{align}\label{gdecay}
\Gamma_{\widetilde{X}\longrightarrow\psi_\mu \gamma} \simeq \dfrac{\text{cos}^2 \theta_\text{W}}{48 \pi M_*^2}\dfrac{m_{\widetilde{X}}^5}{m_{3/2}^2},
\end{align}
where $\theta_\text{W}$ is the electroweak mixing angle. For the NLSP lifetime, $\tau_{\tilde{X}} \lesssim 1$ sec, \cref{gdecay} yields a lower bound on $m_{3/2}$,
\begin{align}
 \left(\dfrac{m_{\widetilde{X}}}{1\, \text{TeV}}\right) \text{GeV} \lesssim m_{3/2} < m_{\tilde{X}}.
\end{align}
For $m_{\tilde{X}}\sim 1$~TeV ($10$~TeV), we obtain $1 \text{ GeV} \lesssim m_{3/2} < 1$~TeV ($10 \text{ GeV} \lesssim m_{3/2} < 10$~TeV). This again confirms that the region bounded by the gray dot-dashed and dashed curves is consistent with a gravitino DM and successful leptogenesis. 

In order to suppress the washout effects in non-thermal leptogenesis we usually require $M_N$ to be somewhat larger than $ T_R$. This can be achieved in the present model by exploiting the freedom in the allowed range of the CP-violating phase factor, $\delta_{eff}\leq 1$. We can choose any value of the lightest right-handed neutrino mass, $M_N$, lying in the range, $T_R\leq M_N \leq m_{\text{inf}}/2$. For example with $M_N = m_{\text{inf}}/4$ we obtain $10\lesssim M_N/T_R \lesssim 200$ consistent with gravitino DM scenario.

\section{\large{\bf Unstable Gravitinos}}

We next consider the following two possibilities for unstable gravitinos:
\begin{enumerate}
	\item Unstable long lived gravitino, $m_{3/2} < 25$ TeV, 
	\item Unstable short lived gravitino, $m_{3/2} > 25$ TeV.
\end{enumerate}
For the unstable long-lived gravitino, $m_{3/2} < 25$ TeV, we have to take into account the BBN bounds on the reheat temperature \cite{Kawasaki:2008qe,Kawasaki:2017bqm},
\begin{align}
 2 \times 10^{7} \, \text{GeV} \lesssim T_R\lesssim ~ 10^{10} \, \text{GeV},
\end{align}
which yields $5\, \text{TeV}\lesssim m_{3/2} \lesssim 25\, \text{TeV}$.
For example, for a typical gravitino mass $\sim 10$~TeV, the BBN bound on the reheat temperature, $T_R \lesssim (1-2) \times 10^9\, \text{GeV}$, is consistent with the predictions displayed in  \cref{fig:reheatr,fig:reheatr2,fig:reheatr3}. Therefore, an unstable long-lived gravitino is viable for a wide range of reheat temperature described above.
\par
For an unstable short lived gravitino, $m_{3/2} > 25$ TeV, the BBN bounds on $T_R$ are no more applicable. However, there is another constraint coming from the decay of gravitinos to the lightest sypersymmetric particle (LSP) $\tilde{\chi}_1^0$. In this case, the LSP relic density is given by, \cite{Kawasaki:2008qe}
\begin{align}\label{neut}
\Omega_{\tilde{\chi}_1^0} h^2 \simeq 2.8 \times 10^{10} \times Y_{3/2}\left(\dfrac{m_{\tilde{\chi}_1^0}}{100\, \text{GeV}}\right),
 \end{align}  
where $m_{\chi_1^0}$ is the LSP mass, and $Y_{3/2}$ is the gravitino yield given as, 
\begin{align}\label{yeild}
Y_{3/2}\simeq 2.3\times 10^{-18}\left(\dfrac{T_R}{10\, \text{TeV}}\right).
\end{align}
Requiring the LSP neutralino density to not exceed the DM relic density, \cref{neut} gives an upper bound on the neutralino mass,
\begin{align}\label{neutr}
m_{\tilde{\chi}_0^1}\lesssim 180\left( \dfrac{10^{10}\, \text{GeV}}{T_R}\right) \text{GeV},\end{align}
which is consistent with the lower limit on the neutralino mass $m_{\tilde{\chi}_0^1}\gtrsim 18$ GeV \cite{Hooper:2002nq}.
Using \cref{neutr}, the predicted range of $T_R$ with successful leptogenesis (i.e., $10^{10} \text{ GeV} \gtrsim T_R \gtrsim \, 2 \times 10^7 \text{GeV}$) can now be translated into a viable mass range for the LSP neutralino,
\begin{align}
180\, \text{GeV}\lesssim m_{\tilde{\chi}_0^1}\lesssim 90\, \text{TeV}.
\end{align}
Thus, an unstable short-lived gravitino scenario is feasible for a wide range of LSP neutralino mass. This is in contrast to earlier studies of $\mu$-hybrid inflation with $M_S \sim 0$ \cite{Rehman:2017gkm,Lazarides:2020zof} where the feasibility of this scenario relies on a non-minimal K\"ahler potential.
\begin{figure*}[t]
	\centering
	\subfloat[\label{gz}]
	{{\includegraphics[width=0.95\columnwidth]{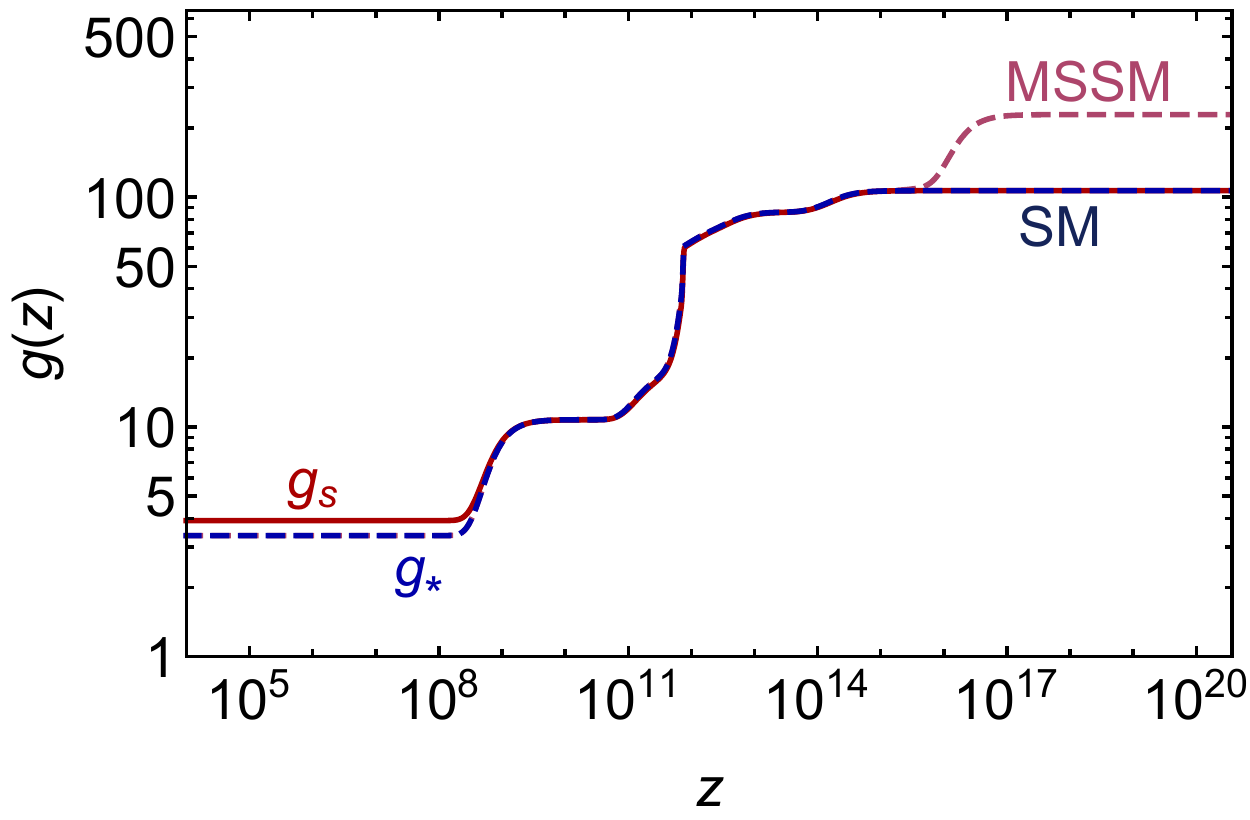} }}%
	\quad
	\subfloat[\label{Gz}]
	{{ 	\includegraphics[width=0.93\columnwidth]{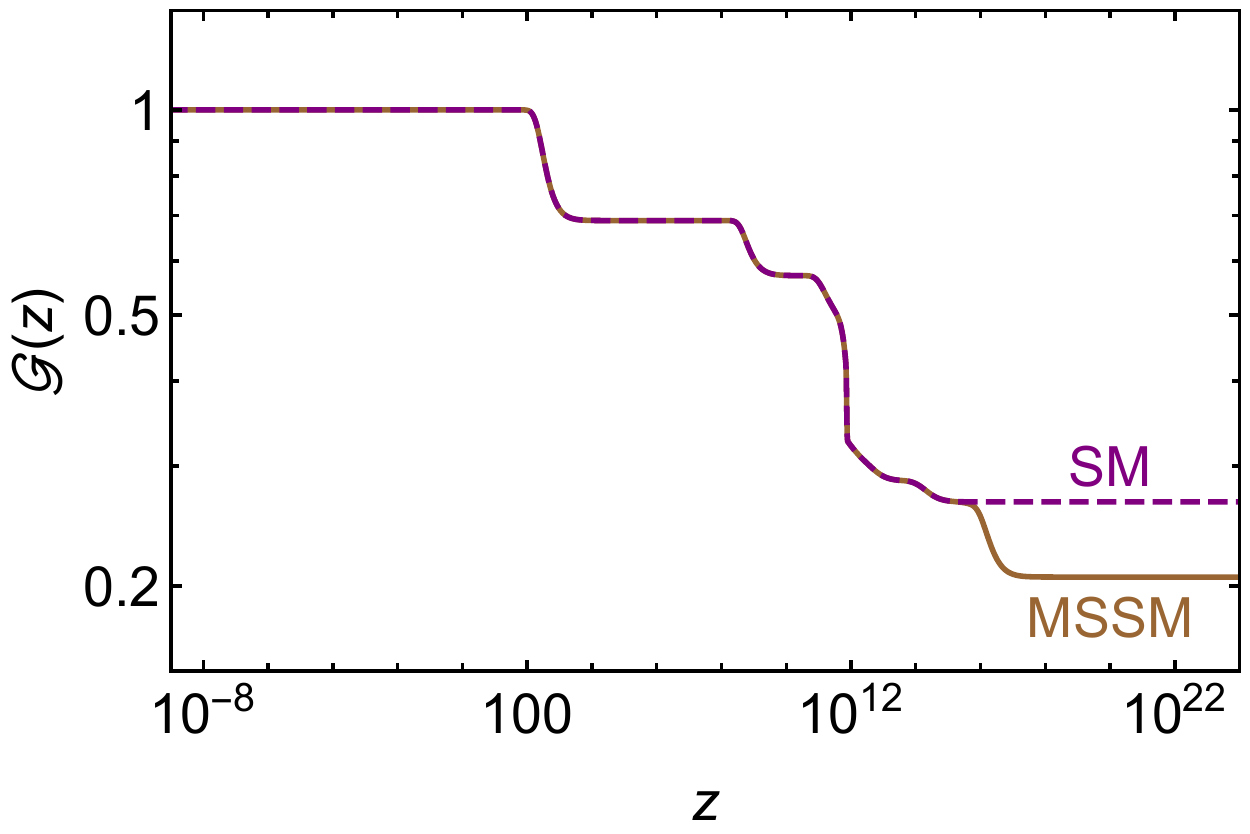} }}%
	\caption{\ref{gz} shows the evolution of effective degrees of freedom with redshift and \ref{Gz} shows the evolution of $\mathcal{G}$ as a function of redshift for both SM and MSSM.}
	\label{fig:effective-degrees-of-freedom-vs}
\end{figure*}
\section{\large{\bf GUT Embedding and Metastable Cosmic Strings}}\label{csdis}
In a SUSY $SO(10)$ framework, a metastable cosmic string network could arise via the symmetry breaking chain,
\begin{equation} \label{so10brk}
SO(10)\longrightarrow \mathbb{G}_{SM}\times U(1)_\chi \longrightarrow \mathbb{G}_{SM},
\end{equation}
where $\mathbb{G}_{SM} \equiv SU(3)_c \times SU(2)_L\times U(1)_Y  $ is the SM gauge group.
The first breaking yields monopoles carrying SM and $U(1)_{\chi}$ magnetic charges. The second breaking yields CSs with the string tension determined by the $U(1)_\chi$ symmetry breaking scale \cite{Buchmuller:2019gfy,Masoud:2021prr}. Assuming that the monopoles are inflated away, the resulting string  network is effectively metastable and yields a stochastic gravitational wave background spectrum that we explore in \cref{SGWB}. Note that the $\chi$ charge coincides with $B-L$ for $Y=0$.
For a recent discussion of metastable CS netwrok formation in other gauge groups see \cite{Buchmuller:2021dtt}.
The metastable string network decays via the Schwinger production of monopole-antimonopole pairs with a rate per string unit length of \cite{Buchmuller:2019gfy,Leblond:2009fq,Monin:2008mp,Monin:2009ch},
\begin{equation}
\Gamma_\text{d}=\frac{\mu_\text{cs}}{2\pi}\text{ exp}(-\pi\kappa_\text{cs}),\quad \kappa_\text{cs}=\frac{m^2}{\mu_\text{cs}},
\end{equation}
where $m\sim M_G$ is the monopole mass and $\kappa_\text{cs}$ quantifies the metastability of CSs network with $\sqrt{\kappa_\text{cs}} \sim 10$ being the stability limit as the lifetime of CSs becomes larger than the age of the Universe. This parameter plays an important role in making predictions for the current and future GW experimental tests.
\section{\large{\bf Stochastic Gravitational Wave Background from Metastable Cosmic Strings}}\label{SGWB}

The stochastic gravitational wave background (SGWB) emitted from the CS network is calculated in terms of the fractional energy density in GWs per logarithmic interval of frequency \cite{Blanco-Pillado:2017oxo},
\begin{equation}\label{3}
\Omega_\text{GW}(f)=\frac{8 \pi G}{3H_{0}^{2}}f (G\mu_\text{cs})^2
\sum_{\text{n}=1}^{\infty} C_\text{n}(f)P_\text{n}.
\end{equation}
Here, $H_0=100 h$ km/s/Mpc is the Hubble parameter today with $h=0.68$ \cite{Planck:2018vyg}, and $P_\text{n}$ is the power spectrum of GWs emitted by the $\text{n}^\text{th}$ harmonic of a CS loop. Our predictions are based on the Blanco-Pillado-Olum-Shlaer (BOS) model \cite{Blanco-Pillado:2017oxo,Blanco-Pillado:2013qja} and cusps as the main source of GWs with $P_\text{n}$ given by \cite{Auclair:2019wcv},
\begin{equation}
P_\text{n}\simeq\frac{\Gamma}{\zeta[\frac{4}{3}]}\text{n}^{-4/3},
\end{equation} 
where $\Gamma \simeq 50$ is a numerical factor specifying the CSs decay rate and $\zeta$ is the Riemann zeta function. It is convenient to work in terms of redshift $z$ with $1+z\equiv a_0/a(t)$ written in terms of the scale factor $a(t)$ and its present value, $a_0$. The number of loops emitting GWs, observed at a given frequency $f$ is defined as \cite{Blanco-Pillado:2017oxo},
\begin{equation}\label{5}
C_\text{n}(f)=\frac{2\text{n}}{f^2}\int_{z_{min}}^{z_{max}}\frac{dz}{H(z)(1+z)^6}{\mathcal{N}}\left(\ell,t\right).
\end{equation}
The integration range corresponds to the life time of CSs network, from it's formation at $z_\text{max}\simeq\frac{T_R}{2.7 \text{K}}$\footnote{$T_R$ is taken to be around $10^{9}\text{ GeV}$.}
  until it's decay at $z_\text{min}=\left(\frac{70}{H_0}\right)^{1/2}\left(\frac{\Gamma (G\mu_\text{cs})^2}{2\pi\times 6.7\times10^{-39}}\text{ exp}(-\pi\kappa_\text{cs})\right)^{1/4}$ \cite{Buchmuller:2020lbh}, and ${\mathcal{N}}\left(\ell,t\right)$ is the number density of CS loop of length $\ell =\frac{2\text{ n}}{(1+z)f}$.

The loop density is defined by considering their formation and decay in different epochs. In a radiation dominated era the loop density is given by \cite{Auclair:2019wcv},
\begin{equation}
{\mathcal{N}}_r(\ell,t)=\frac{0.18}{t^{2/3}(\ell+\Gamma G\mu_\text{cs} t)^{5/2}},
\end{equation}
with $\ell\leq 0.1\,t$, whereas in the matter era it is given by \cite{Auclair:2019wcv},
\begin{equation}
{\mathcal{N}}_m(\ell,t)=\frac{0.27-0.45(\ell/t)^{0.31}}{t^2(\ell+\Gamma G\mu_\text{cs} t)^{2}},
\end{equation}
with $\ell< 0.18 t.$ Lastly, the number density of loops produced during the radiation era, but radiating during the matter era is given by \cite{Auclair:2019wcv}, 
\begin{equation}
{\mathcal{N}}_{r,m}(\ell,t)=\frac{0.18(2H_0\sqrt{\Omega_{r,0}})^{3/2}}{(\ell+\Gamma G\mu_\text{cs} t)^{5/2}}(1+z)^3,
\end{equation}
with $\ell < 0.09 t_\text{eq} - \Gamma \text{G} \mu_\text{cs} t$.\par 

\begin{figure}[t]
		{{\includegraphics[width=\columnwidth]{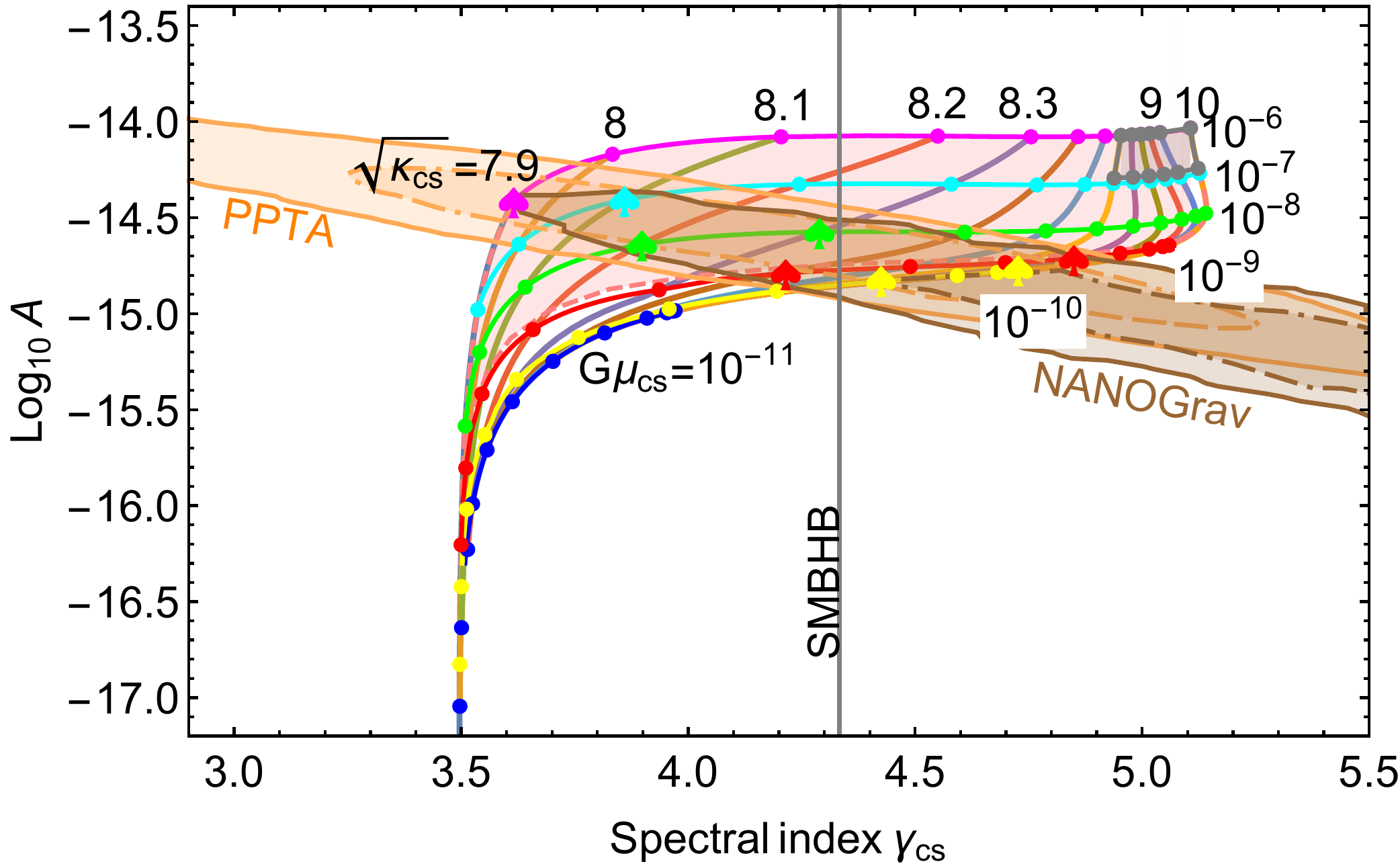}\label{fig:a}}}%
		\centering
	\caption{The GWs signal from stable and metastable CSs compared to the NANOGrav observation for different values of $G\mu_\text{cs}$ and $\kappa_\text{cs}$. Moving from top to bottom $G\mu_\text{cs}$ varies between $10^{-6}$ to $10^{-11}$, while moving from left to right $\sqrt{\kappa_\text{cs}}$ varies between $7.5$ to $9$ and $10$. The gray window in the upper right corner is the region excluded by CMB constraints, and the pink shaded region is for successful leptogenesis, DM and inflation. The brown/orange region, solid (dot-dashed), shows the $2\sigma$ ($1\sigma$) NANOGrav/PPTA posterior contours. The gray vertical line at $\gamma_\text{cs} = 13/3$ represents the slope expected for supermassive black hole binary (SMBHB) mergers.}
	\label{fig:gammaa-plane}
\end{figure}
The cosmological time as a function of $z$ is written as \cite{Blanco-Pillado:2017oxo},
\begin{equation}\label{10}
t(z)=\int_{z}^{+\infty}\frac{dz^\prime}{H(z^\prime)(1+z^\prime)},
\end{equation} 
and the Hubble rate at redshift $z$ is given by \cite{Blanco-Pillado:2017oxo},
\begin{equation}
H(z)=H_0\sqrt{\Omega_{\Lambda,0}+\Omega_{m,0}(1+z)^3+\Omega_{r,0}\mathcal{G}(z)(1+z)^4},
\end{equation}
where $\Omega_{m,0}=0.31$, $\Omega_{r,0}=\frac{4.15\times 10^{-5}}{h^2}$ and $\Omega_{\Lambda,0}=1-\Omega_{m,0}$ are the present values of matter, radiation and dark energy densities respectively, obtained from a standard flat $\Lambda \text{CDM}$ model \cite{Planck:2018vyg}. The function $\mathcal{G}(z)$ defines the change in the expansion rate of the Universe due to  annihilation of relativistic species at earlier times and is given as \cite{Binetruy:2012ze},
\begin{equation}
\mathcal{G}(z)=\frac{g_*(z)g_S^{4/3}(0)}{g_*(0)g_S^{4/3}(z)},
\end{equation}
where $g_*(z)$, $g_S(z)$ are the effective numbers of relativistic and entropic degrees of freedom respectively, at redshift $z$, and $g_*(0)$,  $g_S(0)$ are their present values. The evolution of the effective degrees of freedom with redshift is shown in \cref{gz,Gz}, both for the SM and MSSM\footnote{We thanks Thomas Coleman for sharing the code.}.

Recently, NANOGrav has presented their $12.5$-year data set \cite{NANOGrav:2020bcs} as a characteristic strain of the form,
\begin{equation}
h_{\text{strain}}(f)=A\left(\dfrac{f}{f_\text{yr}}\right)^{(3-\gamma_\text{cs})/2},
\end{equation}
where $f_\text{yr}\equiv 1 \text{yr}^{-1}=32\times10^{-9}$~Hz, $A$ is the strain amplitude and $\gamma_\text{cs}$ is the slope or the spectral index which is related to the spectral GW energy density as,
\begin{equation}\label{slope}
\Omega_\text{GW}(f)=\dfrac{2\pi^2}{3 H_0^2} f^2 h^2(f)=\Omega_\text{yr}\left(\dfrac{f}{f_{yr}}\right)^{5-\gamma_\text{cs}},
\end{equation}
with $\Omega_\text{yr}\equiv\dfrac{2\pi^2 A^2 f_{yr}^2 }{3 H_0^2}$. Taking the first two frequency bins of NANOGrav, i.e, $f_1=2.45\times10^{-9}\text{Hz}$ and $f_2=4.91\times10^{-9}\text{Hz}$, we obtain \cite{Blanco-Pillado:2021ygr},
\begin{align}
\gamma_\text{cs}&=5-\frac{\text{ln}(\Omega_\text{GW}(f_2)/\Omega_\text{GW}(f_1))}{\text{ln}(2)}
\quad\mathrm{and} \notag\\ 
A&=\sqrt{\frac{3H_0^2\Omega_\text{GW}(f_1)f_{yr}^{3-\gamma_\text{cs}}}{2\pi^2f_1^{5-\gamma_\text{cs}}}}.
\end{align} 
\begin{figure*}[t]
	\centering	
	\subfloat[\label{sub:Omegaf}]
	{{	\includegraphics[width=\columnwidth]{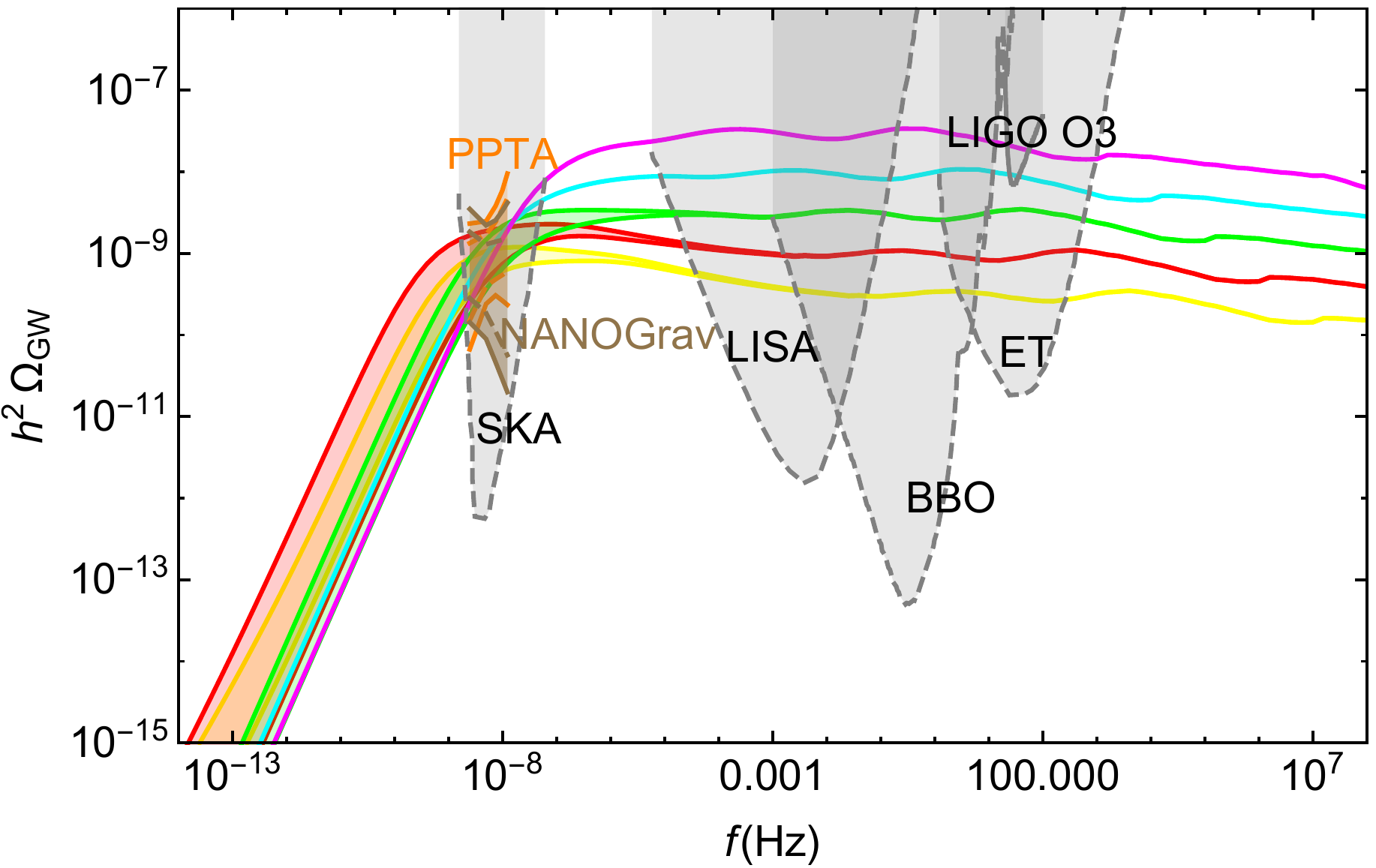}}}%
	\quad
	\subfloat[\label{sub:kappaGmu}]
	{{\includegraphics[width=0.97\columnwidth]{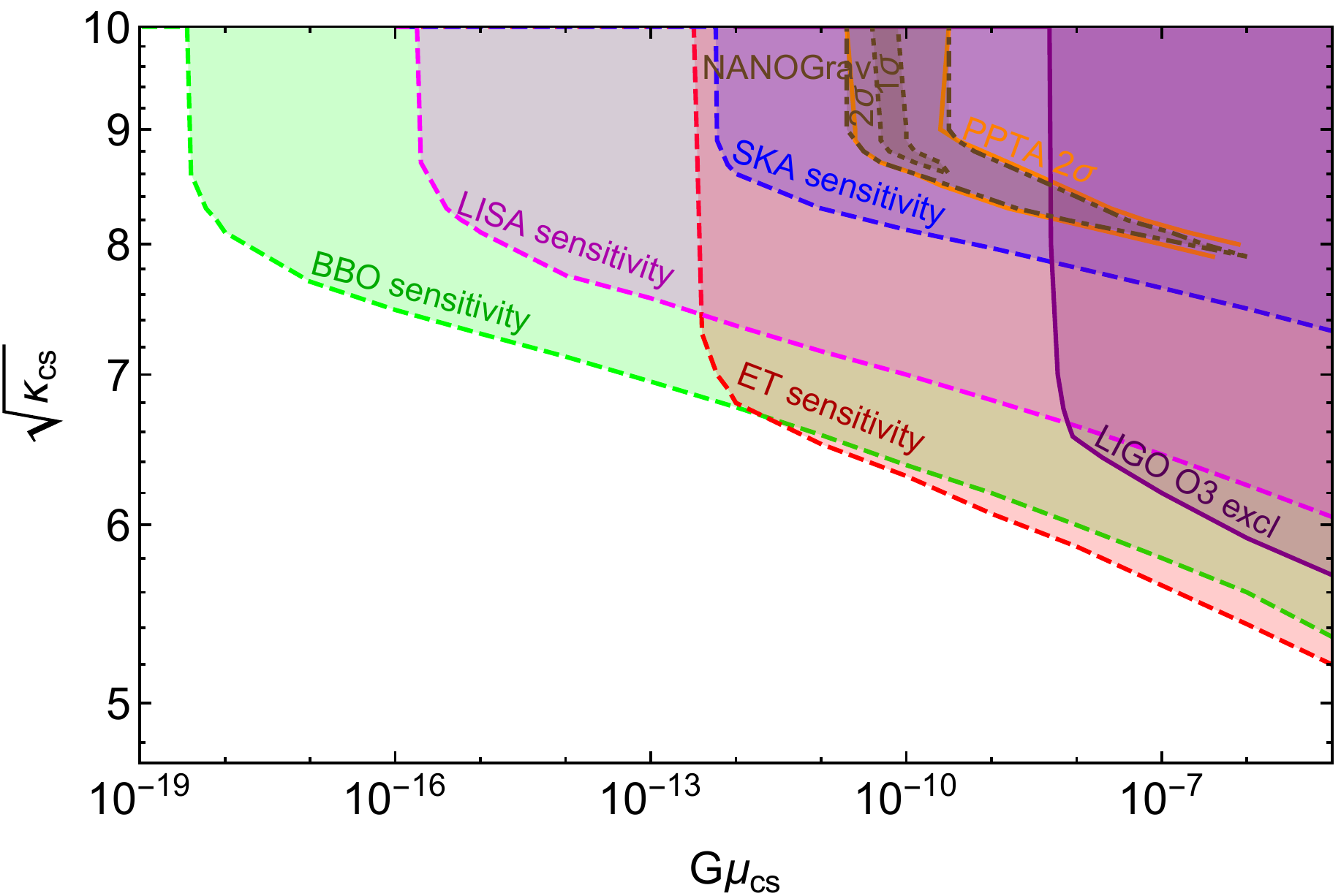}}}%
	\caption{\ref{sub:Omegaf} shows the GW spectra explaining the NANOGrav excess at 2$\sigma$ for allowed values of $G\mu_\text{cs}$ with $\sqrt{\kappa_\text{cs}}=7.9$ to $9$ and $10$ as indicated with corresponding color $\spadesuit$ markers in \cref{fig:gammaa-plane}. The gray shaded regions indicate the sensitivity curves of present (solid boundaries) LIGO O3 \cite{KAGRA:2021kbb} and future (dashed boundaries) SKA \cite{Smits:2008cf}, LISA \cite{amaroseoane2017laser}, ET \cite{Punturo:2010zz}, BBO \cite{Corbin:2005ny} experiments. The brown/orange band solid (dashed) is the NANOGrav/PPTA $2\sigma$ ($1\sigma$). In \ref{sub:kappaGmu}, we display constraints on the CS parameter space from GW experiments. The solid lines are the regions excluded by existing bounds. The dashed line regions indicate the designed sensitivity of future experiments. The dot-dashed (small dashed) region is the NANOGrav excess at $2\sigma$ ($1\sigma$) and the orange region is PPTA 2$\sigma$.}
	\label{fig:kappa-gmu-bounds}
\end{figure*}
This leads to prediction for $(\gamma_\text{cs},A)$, as shown in \cref{fig:gammaa-plane}, with $G\mu_\text{cs}=10^{-6}-10^{-11}$, $\sqrt{\kappa_\text{cs}} \sim 7.5 - 9$, and $10$. Here the parameter $\kappa_\text{cs}$ increases from left to right with $\sqrt{\kappa_\text{cs}}=10$ representing the stable CS limit and $G\mu_\text{cs}$ decreases from top to bottom. The NANOGrav/PPTA \cite{NANOGrav:2020bcs,Goncharov:2021oub}  $2\sigma$ ($1\sigma$) posterior contours are shown by solid (dot-dashed) brown/orange region with a broken power law fit. The gray window in the upper right corner is the region excluded by the CMB constraint and is only applicable to CSs with lifetime longer than CMB decoupling, i.e., $\sqrt{\kappa_\text{cs}}\geq8.6$. The pink shaded region representing successful leptogenesis, DM and inflation is congruous with the gray bounded region of \cref{fig:reheatr,fig:reheatr2,fig:reheatr3}.

In \cref{sub:Omegaf} the spectra of GW are shown for values of $\sqrt{\kappa_\text{cs}}$ and $G\mu_\text{cs}$ which lie within the $2\sigma$ region of NANOGrav as indicated with corresponding color $\spadesuit$ markers in \cref{fig:gammaa-plane}. 
Ignoring dependence on the effective degrees of freedom, the behavior $\Omega_\text{GW}$ $\propto f^{3/2}$ ($\Omega_\text{GW}$ $\propto f^{0}$) is achieved at the low (high) frequency range for the GW spectrum. This corresponds to the range $\gamma_\text{cs} \sim 3.5-5$  via \cref{slope} while explaining most of the predicted region shown in \cref{fig:gammaa-plane}. The CS loops produced during the matter era and loops produced during the radiation era, but radiating during the matter era, become somewhat important in the low frequency range for $\gamma_\text{cs} \gtrsim 5$, but this region lies outside of the NANOGrav bounds. Due to pair production of GUT monopoles, the metastable long strings on superhorizon scales and string loops and segments on subhorizon scales cause a SGWB which we have not considered here but can be seen in \cite{Buchmuller:2021mbb}.

For a detailed comparison of the various present (solid) and future (dashed) experiments, the allowed values of $\sqrt{\kappa_\text{cs}}$ and $G\mu_\text{cs}$
are shown in \cref{sub:kappaGmu}. It is interesting to note that metastable CS with $\sqrt{\kappa_\text{cs}} \sim 8-9 $ allow $G\mu_\text{cs} \sim 10^{-9}-10^{-6}$. Therefore, the gravitino DM scenario with successful leptogenesis in $\mu$-hybrid inflation, combined with NANOGrav SGWB, leads to the predicted range $M_G \sim 10 M \sim 10^{16}\text{ GeV}- 10^{17}\text{ GeV}$ for $G\mu_\text{cs} \sim 10^{-9}-10^{-6}$. However, at larger frequencies, the range $G\mu_\text{cs} \sim 10^{-8}-10^{-6}$ is in some tension with the latest bounds from LIGO O3 \cite{KAGRA:2021kbb}. But this tension still involves some theoretical and experimental uncertainties. A non-standard thermal history \cite{Auclair:2019wcv,Chang:2021afa,Cui:2017ufi,Gouttenoire:2019kij,Gouttenoire:2021jhk}, or late production of the CSs \cite{Lazarides:2021uxv} can ameliorate this tension. Besides NANOGrav, we also show in \cref{sub:kappaGmu} the observable region lying within the sensitivity bounds of future experiments, such as  Einstein Telescope \cite{Punturo:2010zz} and LISA \cite{amaroseoane2017laser}.
\section{\large{\bf Conclusion}}
We have explored the $\mu$-hybrid inflation in a $U(1)_{B-L}$ extension of the MSSM by considering both the linear and quadratic soft SUSY breaking terms with special focus on the parameter space described by $|M_S| \gg  m_{3/2}$. A wide range of the gauge symmetry breaking scale, $6 \times 10^{14}\lesssim M/\text{GeV}\lesssim10^{16}$ is predicted for successful non-thermal leptogenesis and stable gravitino as a viable dark matter candidate. This parameter range corresponds to a stochastic gravitational wave background from a metastable cosmic string network with tension $G\mu_\text{cs}\sim 10^{-9}-10^{-6}$. Such a metastable cosmic string network can arise in a grand unified theory with $U(1)_{B-L}$ embedded in an $SO(10)$ model. An order of magnitude splitting is predicted between the GUT and $B-L$ breaking scales.
 This connection certainly provides a unique opportunity to probe the seesaw mechanism and leptogenesis with gravitational waves \cite{Dror:2019syi,Blasi:2020wpy,Samanta:2020cdk,Samanta:2021zzk}. The metastable cosmic string network lies within the 2$\sigma$ NANOGrav 12.5 year data/PPTA and is also within reach of future gravitational wave experiments.


\section*{\large{\bf Acknowledgments}}
This work is partially supported by the DOE grant No. DE-SC0013880 (Q.S.). Adeela
Afzal thanks Kai Schmitz, Valerie Domcke, Wilfried Buchmuller, Ken D. Olum and George Lazarides for
useful discussion. 

\bibliographystyle{unsrt}
\bibliography{SUSYhybrid}	  
	 
\end{document}